%% file: draft.tex
\documentclass[12pt,a4paper]{article}

\usepackage{amssymb,amsmath,amsfonts,amsthm, amscd, mathrsfs,helvet, mathtools}
\usepackage{graphicx,verbatim}
\usepackage[usenames, dvipsnames]{color}
\usepackage{psfrag}
\usepackage{cite}
\usepackage{srcltx}

\theoremstyle{plain}

\newtheorem*{theorem*}{Theorem}

\newtheorem*{proposition*}{Proposition}

\newcommand{\tensor}[1]{{\bf \underline{#1}}}

\definecolor{brightBlue}{rgb}{0,0,1}

\definecolor{Violet}{rgb}{0.47,0,1}

\DeclareMathOperator{\tr}{tr}

\def\b{\mathfrak{b}}

\def\f{\mathfrak{f}}
\def\g{\mathfrak{g}}
\def\h{\mathfrak{h}}

\def\n{\mathfrak{n}}

\def\hf{\widehat{\mathfrak{f}}}

\def\ha{\mbox{\small $\frac{1}{2}$}}
\def\qa{\mbox{\small $\frac{1}{4}$}}

\def\CC{\mathbb{C}}

\def\P{\mathcal{P}}
\def\R{\mathcal{R}}
\def\L{\mathcal{L}}

\def\1{\tensor{1}}
\def\2{\tensor{2}}
\def\3{\tensor{3}}
\def\4{\tensor{4}}

\def\beq{\begin{equation}}
\def\eeq{\end{equation}}
\def\beqz{\begin{equation*}}
\def\eeqz{\end{equation*}}
\def\bea{\begin{eqnarray}}
\def\eea{\end{eqnarray}}

\def\pcm{{principal chiral model }}
\def\pbs{{Poisson brackets }}
\def\pb{{Poisson bracket }}
\def\Klimcik{{Klim$\check{\text{c}}$\'{\i}k} }
\def\dss{ {\delta_{\sigma\sigma'}} }

\def\l{l}
\def\AA{\L^g}
\def\QQ{\mathfrak{J}}
\def\pp{\mathsf{p}}
\def\qq{\mathsf{q}}

\numberwithin{equation}{section}

\begin{document}

\begin{center}
\vspace*{2em}
{\large\bf
On classical $q$-deformations of integrable $\sigma$-models}\\
\vspace{1.5em}
F. Delduc$\,{}^1$, M. Magro$\,{}^1$, B. Vicedo$\,{}^2$

\vspace{1em}
\begingroup\itshape
{\it 1) Laboratoire de Physique, ENS Lyon
et CNRS UMR 5672, Universit\'e de Lyon,}\\
{\it 46, all\'ee d'Italie, 69364 LYON Cedex 07, France}\\
\vspace{1em}
{\it 2) School of Physics, Astronomy and Mathematics,
University of Hertfordshire,}\\
{\it College Lane,
Hatfield AL10 9AB,
United Kingdom}
\par\endgroup
\vspace{1em}
\begingroup\ttfamily
Francois.Delduc@ens-lyon.fr, Marc.Magro@ens-lyon.fr, Benoit.Vicedo@gmail.com
\par\endgroup
\vspace{1.5em}
\end{center}

\paragraph{Abstract.}
A procedure is developed for constructing deformations of integrable $\sigma$-models
which are themselves classically integrable. When applied to the principal
chiral model on any compact Lie group $F$, one recovers the Yang-Baxter $\sigma$-model
introduced a few years ago by C. {Klim$\check{\text{c}}$\'{\i}k}. In the case of the symmetric space
$\sigma$-model on $F/G$ we obtain a new one-parameter family of
integrable $\sigma$-models.
The actions of these models correspond to a deformation of
the target space geometry and include a torsion term.
An interesting feature of the construction is the $q$-deformation of the symmetry corresponding to left multiplication in the original models, which becomes replaced by a classical $q$-deformed Poisson-Hopf algebra. 
Another noteworthy aspect of the deformation in the coset $\sigma$-model case is that it interpolates between a compact and a non-compact symmetric space. This is exemplified in the case of the $SU(2)/U(1)$ coset $\sigma$-model which interpolates all the way to the $SU(1,1)/U(1)$ coset $\sigma$-model.

\section{Introduction}

The property of integrability is extremely scarce among two-dimensional $\sigma$-models.
And yet when present it provides a powerful tool in the study of various exact properties of these
models.
There is, however, no systematic way of proving whether or not a two-dimensional $\sigma$-model is integrable.
In light of this, an interesting question to consider is the following: given an integrable $\sigma$-model, is it possible to construct a deformation of this model which is itself integrable?

\medskip

In the case of the $SU(2)$ principal chiral model, an example of such a deformation is given by
the diagonal anisotropic $SU(2)$ principal chiral model introduced by Cherednik in
\cite{Cherednik:1981df}. The action for the $SU(2)$-valued field $g$ of this model
may be written as
\begin{equation*}
S_{\rm C}[g] = - \ha \int d\tau d\sigma \, \tr \left( \text{ad}(\partial_+ g \, g^{-1}) \, J \,
 \text{ad}(\partial_- g \, g^{-1}) \right),
\end{equation*}
where $J = \text{diag}(J_1, J_2, J_3)$ is a diagonal matrix, the effect of which is to deform
the metric away from the Killing form of $\mathfrak{su}(2)$. This model is known to be integrable
\cite{Cherednik:1981df} and provides a two-parameter deformation of the
principal chiral model.

In the special case $J_1 = J_2 \neq J_3$ it reduces to
the squashed sphere $\sigma$-model, where the parameter $C = J_3/J_1$
describes the squashing of the $3$-sphere. As a result of this squashing
 when $C \neq 1$, the global $SU(2)_L
\times SU(2)_R$ symmetry of the principal chiral model is broken down to $SU(2)_L \times
U(1)_R$. However, it was recently argued in \cite{Kawaguchi:2011pf, Kawaguchi:2012gp}
that a certain deformation of the $SU(2)_R$ symmetry is still realised in the squashed
sphere $\sigma$-model. Specifically, as the deformation is turned on, the $SU(2)_R$
symmetry gets replaced by a classical $q$-deformed $U^{\P}_q(\mathfrak{sl}_2)$ symmetry,
where the algebraic deformation parameter $q = q(C)$ is a function of the geometric
squashing parameter $C$.

\medskip

A generalisation of the above one-parameter deformation for the principal chiral model on any
compact Lie group $F$ is the so called Yang-Baxter $\sigma$-model introduced by \Klimcik  in \cite{Klimcik:2002zj}. In a subsequent
paper it was then proved that this deformation is in fact also integrable \cite{Klimcik:2008eq}.
Using the conventions of the present paper, the action of this model reads
\begin{equation*}
S_{\rm K}[g] =
-\ha \int d\tau d\sigma \, \kappa \left( \partial_+ g \, g^{-1},
\frac{(1 + \eta^2)^2}{1 - \eta R} \partial_- g \, g^{-1} \right),
\end{equation*}
where $\kappa$ is the Killing form of the Lie algebra $\f = \text{Lie}(F)$ and
$\eta \geq 0$ is the deformation parameter. Here $R$ is a certain solution of
the modified classical Yang-Baxter equation on $\f$. In the limit $\eta \to 0$ this action
reduces to that of the principal chiral model. Furthermore, in the case $F = SU(2)$
it reduces to the action of the diagonal anisotropic $SU(2)$ principal chiral model
with $J_1 = J_2 \neq J_3$.

\medskip

The first objective of this paper is to put forward a procedure for
deforming integrable $\sigma$-models in a way which manifestly preserves their integrability. The cases that we shall consider here are the principal chiral model on any compact Lie group $F$ and the coset $\sigma$-model on a symmetric space $F/G$. The second objective is to show that the models so obtained admit a classical $q$-deformed symmetry.  

\medskip

In the case of the principal chiral model, we shall in fact recover in this way
the Yang-Baxter $\sigma$-model. Its integrability will, however, be
automatic from our construction. Furthermore,
working in the Hamiltonian formalism will also enable us to
show that the Yang-Baxter $\sigma$-model admits a classical $q$-deformed $U^{\P}_q(\f)
\times F_R$ symmetry, where $q = q(\eta)$ is a certain function of the deformation parameter
$\eta$. In the limit $\eta \to 0$ this reduces to the global $F_L \times F_R$ symmetry of the
principal chiral model. This feature of the Yang-Baxter $\sigma$-model therefore generalises
the analogous $q$-deformation exhibited in \cite{Kawaguchi:2011pf, Kawaguchi:2012gp} for
the symmetries of the squashed sphere $\sigma$-model.

\medskip

Most importantly, our procedure admits a straightforward generalisation to coset
$\sigma$-models. We shall indeed construct a new one-parameter deformation of the coset
$\sigma$-model on $F/G$ where $F$ is a compact Lie group and $G = \exp \g$ is the
Lie group associated with the subalgebra $\g$ of $\f$ fixed by an order $2$ automorphism
$\sigma : \f \to \f$. The resulting action takes the form
\begin{equation*}
S[g] = - \ha \int d\tau d\sigma \, \kappa\biggl((g^{-1} \partial_+g)^{(1)}, \frac{1+\eta^2}{1-\eta   R_g \circ P_1} (g^{-1} \partial_- g)^{(1)} \biggr),
\end{equation*}
where $R_g = \text{Ad}\, g^{-1} \circ R \circ  \text{Ad}\, g$ and
$P_1 M = M^{(1)}$ is the projection of $M \in \f$   onto the subspace of $\f$ on which the
automorphism $\sigma$ has eigenvalue $-1$. Just as in the case of the Yang-Baxter
$\sigma$-model, we will show that this model also admits a $q$-deformed $U^{\P}_q(\f)$
symmetry where $q = q(\eta)$ is again a function of the real deformation parameter $\eta$.

\medskip
 
Our strategy for deforming the principal chiral model and coset $\sigma$-models
crucially exploits the existence  
of a second Poisson bracket compatible with the original one. Such a compatible bracket was introduced in \cite{Faddeev:1985qu} for the $SU(2)$ principal chiral model and this was subsequently generalised to all other principal chiral models and coset $\sigma$-models in \cite{Delduc:2012qb}.
Recall that the integrability of these models at the Hamiltonian level follows from the Poisson bracket of their Lax matrix taking the specific form in \cite{Maillet:1985fn, Maillet:1985ek}. In order to construct an integrable deformation we should therefore ensure that this latter property is preserved.
Now in both models, the Lax matrix depends on the
canonical fields only indirectly through certain currents.
 We shall not modify this dependence of the Lax matrix on these currents. Instead, what we shall deform is the way these currents depend on the underlying canonical fields. This will be achieved by deforming the Poisson bracket of the currents, which we do by adding a multiple of the compatible Poisson bracket. As a result, the Hamilton dynamics of the canonical fields will be deformed.
After taking the inverse Legendre transform this procedure leads to the above
Lagrangians for the deformed models.

   \medskip

This article is organised as follows.
The procedure is first presented in the case
of the principal chiral model in section \ref{secpcmdefo}. After recalling some
well known properties relating to the integrability and symmetries of this model, we introduce
the deformed Poisson bracket in subsection \ref{secpcmdsetting}.
The resulting deformation of the relation between
the Lax matrix and canonical variables is worked out in subsections \ref{june3ad} and
\ref{secpcm24}. The deformation of the global $F_L \times F_R$ symmetry is studied
in the next subsection. We end this section by deriving
the action describing our deformed model, thereby showing that it coincides
with the Yang-Baxter $\sigma$-model.
Section \ref{secdeformedcoset} is devoted to the deformation
of symmetric space $\sigma$-models. We follow exactly the same steps as for the principal chiral model.
The corresponding action is computed in subsection \ref{secggyb}. In section
\ref{secsu2u1} we study the simplest example of the deformed $SU(2)/U(1)$ coset $\sigma$-model.
It provides an interesting interpolation between coset $\sigma$-models on
the compact and non-compact symmetric spaces $SU(2)/U(1)$ and $SU(1,1)/U(1)$, respectively.
This article includes four appendices. Some notations on compact real Lie algebras and
a reminder on the Iwasawa decomposition are found in Appendix \ref{app: real form}.
Details for the proof of the $q$-Poisson-Serre relations are given in Appendix
\ref{app: q-Serre}. Finally, the last two appendices are respectively devoted to
a discussion of the modified classical Yang-Baxter equation and the deformed
Poisson bracket used in the case of the coset $\sigma$-models.

\section{Deforming the principal chiral model}
\label{secpcmdefo}

\subsection{Principal chiral model} \label{sec: pcm}

We begin this section by reviewing aspects of the principal chiral model on a compact Lie
group $F$ which will be relevant for our purposes. Although these are standard properties,
it is important to recall them in order to emphasise those features of the model which we
shall deform later.

\paragraph{Hamiltonian, equations of motion and Lax matrix.}

The principal chiral model may be described by a pair of fields
$j_0(\sigma)$ and $j_1(\sigma)$ each of which takes values in the compact Lie algebra
$\f = \text{Lie}(F)$. We shall consider the case where the underlying space, parameterised
by $\sigma$, is the entire real line. In particular, the fields $j_0(\sigma)$ and $j_1(\sigma)$
will be assumed to decay sufficiently rapidly at infinity. Their \pbs are given by
\begin{subequations} \label{orpcm}
\begin{align}
\label{orpba} \{ j_{0\1}(\sigma), j_{0\2}(\sigma') \}  &=   -[C_{\1\2}, j_{0\2}(\sigma)] \delta_{\sigma \sigma'}\\
\label{orpbb} \{ j_{0\1}(\sigma), j_{1\2}(\sigma') \}  &=  - [C_{\1\2}, j_{1\2}(\sigma)] \delta_{\sigma \sigma'} + C_{\1\2} \delta'_{\sigma \sigma'}\\
\label{orpbc} \{ j_{1\1}(\sigma), j_{1\2}(\sigma') \}  &= 0.
\end{align}
\end{subequations}
We denote by $C_{\1\2} = \kappa_{ab} T^a \otimes T^b$ the tensor Casimir with
$\kappa_{ab}$ the components of the inverse of the Killing form $\kappa$ on $\f$ in any basis $T^a$
(see appendix \ref{app: real form} for notations).

The Hamiltonian of the model reads
\begin{equation} \label{H PCM}
H_{\rm PCM} =  -\ha \int_{-\infty}^{+\infty} d\sigma \bigl(\kappa(j_0,j_0)+\kappa(j_1,j_1)\bigr).
\end{equation}
The resulting equations of motion, with $\partial_{\tau} = \{ H_{\rm PCM} ,\cdot\}$, take the form of the conservation equation and the zero curvature equation
\begin{subequations} \label{may30b}
\begin{align}
-\partial_{\tau} j_0 + \partial_{\sigma} j_1 &=0, \label{may30a}\\
\partial_{\tau} j _1 - \partial_{\sigma} j_0 -[j_0,j_1]&=0.
\end{align}
\end{subequations}
The integrability of these equations of motion is encoded in the usual Lax matrix
\begin{equation} \label{Lax PCM}
\L(\lambda) = \frac{1}{1 - \lambda^2} \left( j_1 + \lambda \, j_0 \right),
\end{equation}
which takes values in the loop algebra $\hf = \f \otimes \mathbb{C}(\!( \lambda )\!)$.

\paragraph{Symmetry algebra and group valued field.}

It is instructive to recall some properties of the global $F_L \times F_R$ symmetry of the principal chiral model. Indeed, part of these symmetries will turn out to be deformed in the model we shall construct.

It is immediate from equation \eqref{may30a} that $Q^R = \int d\sigma j_0$
is a conserved quantity. By introducing the group valued principal chiral
field $g \in F$ through the relation $j_1 = - g^{-1} \partial_{\sigma} g$, this charge is seen
to generate the $F_R$
symmetry of the model acting as $g \mapsto g U_R$. Indeed, the \pbs \eqref{orpbb} and
\eqref{orpbc} lifted to the field $g$ read
\begin{subequations} \label{j0gPB}
\begin{align}
\{  j_{0\1}(\sigma),g_{ \2}(\sigma') \}  &=   g_{\2}(\sigma) C_{\1\2}  \dss, \label{may30c} \\
\{ g_{\1}(\sigma),g_{ \2}(\sigma') \}  &= 0. \label{june2a}
\end{align}
\end{subequations}
Furthermore, this charge $Q^R$ appears at order $\lambda^{-1}$ in the expansion of the monodromy
matrix at $\lambda = \infty$ since the expansion of the Lax matrix \eqref{Lax PCM}
there begins with
\begin{equation} \label{Lax at 0}
\L(\lambda) = - \lambda^{-1} j_0 + O(\lambda^{-2}).
\end{equation}

It turns out that both the field $g$ and the $F_L$ symmetry, acting as $g \mapsto U_L g$, may be
conveniently described in terms of the leading behaviour of the Lax matrix at the point $\lambda = 0$.
By virtue of the definition of $j_1$ in terms of $g$, the value of the Lax matrix at $\lambda = 0$ is
$\L(0) = - g^{-1} \partial_{\sigma} g$.  This shows that the field $g \in F$ may be characterised rather
abstractly as the gauge transformation parameter which sends $\L(0)$ to zero.
The generator of the $F_L$ symmetry can then be extracted from the next order in the expansion
of the gauge transformed Lax matrix at $\lambda = 0$. Indeed, if we define   $\l_0 = g j_0 g^{-1}$
we have
\begin{equation} \label{Lax PCM 0}
\AA(\lambda)\coloneqq \partial_\sigma g g^{-1} + g \L(\lambda)  g^{-1}=\lambda  \l_0   + O(\lambda^2).
\end{equation}
Furthermore, the definition of $\l_0$ and the \pbs \eqref{orpba} and \eqref{j0gPB} lead to
\begin{subequations} \label{gl0 PB}
\begin{align}
\label{gl0 PB a} \{ \l_{0 \1}(\sigma), \l_{0 \2}(\sigma') \} &= \big[ C_{\1\2}, \l_{0 \2}(\sigma) \big] \delta_{\sigma \sigma'},\\
\label{gl0 PB b} \{ \l_{0 \1}(\sigma), g_{\2}(\sigma') \} &= C_{\1\2} \, g_{\2}(\sigma) \delta_{\sigma \sigma'}.
\end{align}
\end{subequations}
It therefore follows that the generator of the $F_L$ symmetry is $Q^L = \int d\sigma l_0$ and moreover
it appears as the coefficient of $\lambda$ in the expansion at $\lambda= 0$ of the gauge transformed
monodromy matrix.

Let us briefly summarise the above by remarking that the pair of fields $g$ and $l_0$ may roughly
speaking be regarded as canonical fields for the principal chiral model with \pbs given in
\eqref{june2a} and \eqref{gl0 PB}. The pair $(g, l_0)$ takes values in the canonical right
trivialisation of the cotangent bundle of $F$. Moreover, both these fields may be extracted from the
Lax matrix using the following scheme:
\begin{itemize}
\item The field $g$ is characterised by the condition $\AA(0)=0$ which fixes $ j_1 = - g^{-1} \partial_\sigma g$.
\item   The field $\l_0$ is obtained as $\frac{\partial \AA }{\partial\lambda}(0) = \l_0 $, implying the
relation   $ j_0 =  g^{-1} l_0 g$.
\end{itemize}

\subsection{Setting up the deformation}
\label{secpcmdsetting}
\paragraph{Deformed Poisson bracket.}

Our starting point for constructing a deformation of the principal chiral model in the Hamiltonian
formalism will be to deform its Poisson bracket. A natural way to do this is to combine the original
Poisson bracket $\{ \cdot, \cdot \}$ of the current in \eqref{orpcm} with a compatible Poisson bracket, say $\{ \cdot, \cdot \}'$. In
the case at hand there is a natural candidate for $\{ \cdot, \cdot \}'$, namely the Poisson bracket associated
with the Faddeev-Reshetikhin model \cite{Faddeev:1985qu}. Indeed, its compatibility with \eqref{orpcm}
was shown in \cite{Delduc:2012qb}. We therefore consider the following linear combination of Poisson brackets
\begin{equation} \label{deformed PB PCM}
\{ \cdot, \cdot \}_{\epsilon} \coloneqq \{ \cdot, \cdot \} + \epsilon^2 \{ \cdot, \cdot \}',
\end{equation}
where the parameter $\epsilon$ is taken to be real and positive.
 When $\epsilon = 0$ this bracket corresponds to the original undeformed Poisson bracket $\{ \cdot, \cdot \}_0 = \{ \cdot, \cdot \}$ whereas when $\epsilon$ tends to infinity it becomes proportional to the Faddeev-Reshetikhin bracket $\{ \cdot, \cdot \}'$. For any other value $\epsilon > 0$ it reads
\begin{subequations} \label{interpolating PB PCM}
\begin{align}
\label{interpolating PB a} \{ j_{0\1}(\sigma), j_{0\2}(\sigma') \}_{\epsilon} &= -(1 - \epsilon^2) [C_{\1\2}, j_{0\2}(\sigma)] \delta_{\sigma \sigma'},\\
\label{interpolating PB b} \{ j_{0\1}(\sigma), j_{1\2}(\sigma') \}_{\epsilon} &= -(1 - \epsilon^2) [C_{\1\2}, j_{1\2}(\sigma)] \delta_{\sigma \sigma'} + C_{\1\2} \delta'_{\sigma \sigma'},\\
\label{interpolating PB c} \{ j_{1\1}(\sigma), j_{1\2}(\sigma') \}_{\epsilon} &=  \epsilon^2 [C_{\1\2}, j_{0\2}(\sigma)] \delta_{\sigma \sigma'}.
\end{align}
\end{subequations}

\paragraph{Lax matrix and Hamiltonian.} In order to ensure that the deformed model remains integrable as we vary the deformation parameter $\epsilon$, we shall do two things.

On the one hand, and in the spirit of \cite{Delduc:2012qb}, we shall require that the Lax matrix of the deformed model be the same function of $j_0$ and $j_1$, independent of $\epsilon$. In other words, we will take the same Lax matrix \eqref{Lax PCM} for every value of the parameter $\epsilon$.

On the other hand, we shall also insist that the dynamics of the fields $(j_0, j_1)$ remain the same as we vary $\epsilon$. Nevertheless, since the \pbs of $(j_0,j_1)$ do depend on $\epsilon$, this implies that the dependence of $(j_0,j_1)$ on the canonical fields will vary with $\epsilon$. Consequently, the dynamics of these canonical fields will be deformed. When $\epsilon$ vanishes, the principal chiral field $g$ itself together with the field $j_0$, or equivalently $l_0$, may be regarded as canonical fields in view of \eqref{june2a} and \eqref{gl0 PB}.
The possibility to deform the principal chiral model will therefore come from the freedom in defining the field $g$ at non-zero values of the deformation parameter $\epsilon$.
We shall come back in detail to this important point in section \ref{june3ad} below.

The Lax matrix \eqref{Lax PCM} depends linearly on the fields $(j_0, j_1)$. Therefore, in order to find the Hamiltonian $H^{\epsilon}$ which generates the same dynamics on these fields as the principal chiral model but with respect to the deformed Poisson bracket \eqref{interpolating PB PCM}, we should solve the following equation
\begin{equation} \label{Hc Hpcm}
\{ H^{\epsilon}, \L \}_{\epsilon} = \{ H_{\rm PCM}, \L \}.
\end{equation}
By using the fact that the Hamiltonian $H_{\rm PCM}$ has vanishing Faddeev-Reshetikhin Poisson bracket with any function of $(j_0, j_1)$, it is easy to see that $H^{\epsilon} = H_{\rm PCM}$ is also the Hamiltonian with respect to the deformed bracket.

\paragraph{Deformed twist function.}

In view of deforming the definition of the principal chiral field $g$ as given in section \ref{sec: pcm}, we first need to understand the distinguishing characteristic of the special point $\lambda = 0$ entering this definition.

In the Hamiltonian formalism, the algebraic ingredients underpinning the integrability of non-ultralocal models of interest in this paper were emphasised in \cite{Delduc:2012qb}, to which the reader is referred. Aside from the loop algebra $\hf$ and the Lax matrix $\L(\lambda)$ valued in $\hf$, an essential role is played by the standard split $R$-matrix $\R$, which is a solution of the modified classical Yang-Baxter equation on $\hf$ (see appendix \ref{app: mCYBE}). An equally important ingredient in this setup is the twist function $\varphi(\lambda)$.
As explained in \cite{Delduc:2012qb}, in this language the \pb of any two functions of the Lax matrix may be expressed in terms of the rational inner product on $\hf$ and the twisted $R$-matrix $\R \circ \tilde{\varphi}^{-1}$, where $\tilde{\varphi}$ denotes multiplication by the twist function $\varphi(\lambda)$. The twist functions of the principal chiral model and the Faddeev-Reshetikhin model are given respectively by
\begin{equation} \label{twists PCM FR}
\varphi_{\rm PCM}(\lambda) = -1 + \frac{1}{\lambda^2}, \qquad
\varphi_{\rm FR}(\lambda) = 1.
\end{equation}
Note that in this formalism, the compatibility between the \pbs of these two models may be inferred from \cite{Reyman:1988sf}.

The Poisson bracket $\{ f_1, f_2 \}_{\epsilon}(\L)$ of any two functions $f_1$ and $f_2$ can be computed in two ways. By definition, it is given by the linear combination of the brackets $\{ f_1,f_2 \}(\L)$ and $\{f_1, f_2 \}'(\L)$ which are respectively linear in $\R \circ \tilde{\varphi}^{-1}_{\rm PCM}$ and $\R \circ \tilde{\varphi}^{-1}_{\rm FR}$. Alternatively, one can determine the twist function $\varphi_{\epsilon}$ for the deformed Poisson bracket \eqref{deformed PB PCM} and then compute $\{ f_1, f_2 \}_{\epsilon}(\L)$ directly in terms of $\R_\epsilon \coloneqq \R \circ \tilde{\varphi}^{-1}_{\epsilon}$.
Restricting to linear functions $f_1$ and $f_2$ of $\L$, one has
\beqz
\{ \L_{\1}(\sigma), \L_{\2}(\sigma') \}_\epsilon = [\R_{\epsilon\1\2}, \L_{\1}(\sigma)] \delta_{\sigma \sigma'} - [\R^{\ast}_{\epsilon\1\2}, \L_{\2}(\sigma)] \delta_{\sigma \sigma'} + (\R_{\epsilon\1\2} + \R^{\ast}_{\epsilon\1\2}) \delta'_{\sigma \sigma'}
\eeqz
where $\R_{\epsilon\1\2}$ and $\R^{\ast}_{\epsilon\1\2}$ are respectively the kernels
of $\R_\epsilon$ and its adjoint with respect to the rational inner product on $\hf$. The reader
is referred to \cite{Delduc:2012qb} for details.

Putting all this together we obtain a simple expression for the inverse of the twist function of the deformed Poisson bracket
\begin{equation*}
\varphi_{\epsilon}(\lambda)^{-1} = \varphi_{\rm PCM}(\lambda)^{-1} + \epsilon^2 \varphi_{\rm FR}(\lambda)^{-1}.
\end{equation*}
Substituting the definitions \eqref{twists PCM FR} we find the deformed twist function to be
\begin{equation}
\varphi_{\epsilon}(\lambda) = \frac{1 - \lambda^2}{(1 - \epsilon^2) \lambda^2 + \epsilon^2}.
\label{may31a}
\end{equation}

\paragraph{Poles of the deformed twist function.} It is clear
from \eqref{twists PCM FR} that the point $\lambda=0$, from which the
principal chiral field $g$ may be extracted, corresponds in fact to the pole
of the twist function $\varphi_{\rm PCM}(\lambda)$. It is therefore natural
to expect that the poles of the deformed twist function \eqref{may31a} will
be of particular importance in defining the group valued field corresponding
to the deformed theory. Moreover, the symmetry generators of the deformed
model will be obtained by expanding the monodromy matrix around these points.
They are located at
\begin{equation} \label{twist poles PCM}
\lambda_{\pm} = \pm \frac{i \epsilon}{\sqrt{1 - \epsilon^2}}
\end{equation}
and have the property $\lambda_- = \overline{\lambda}_+$ which we will
make use of later. Hence, the double pole at $\lambda = 0$ of the twist
function $\varphi_{\rm PCM}(\lambda)$ is seen to split into a pair of single
poles as we turn on the deformation parameter
$\epsilon$.
Another interesting feature of \eqref{twist poles PCM} is that the poles move off to infinity as
$\epsilon \to 1$.

\subsection{Defining the group valued field}
\label{june3ad}

\paragraph{Definition of $g$.} Mimicking the interpretation of the principal chiral field as the parameter of a gauge transformation sending the Lax matrix $\L(0)$ to zero, we would like to define the field $g$ for $\epsilon \neq 0$ as the parameter of a gauge transformation of some sort.
However, since for $\epsilon \neq 0$ there are now two poles at $\lambda_{\pm}$,
we should consider both Lax matrices $\L(\lambda_+)$ and $\L(\lambda_-)$.

Consider first $\L(\lambda_+)$.
Since we want the field $g$ to belong to the compact Lie group $F$ for any $\epsilon$, \emph{i.e.} $g^\dagger = g^{-1}$, we should ensure that $\partial_\sigma g g^{-1}$ takes values in $\f$. We therefore define $g$ so that $\partial_\sigma g g^{-1}$ coincides with the component along $\f$ in the Iwasawa decomposition \eqref{Iwasawa} of $- g \L(\lambda_+) g^{-1}$. In other words, we define the field $g \in F$ as the parameter of a gauge transformation such that
\begin{subequations} \label{gauge Lax}
\begin{equation} \label{gauge Lax a}
\AA(\lambda_+) = \partial_{\sigma} g  g^{-1} + g \L(\lambda_+) g^{-1}
\end{equation}
belongs to $\h_0 \oplus \n^+ \subset \b^+$, where $\h_0, \n^+$ and $\b^+$ are defined in appendix \ref{app: real form}. Consider now the effect of this gauge transformation at the other point $\lambda = \lambda_-$, namely
\begin{equation} \label{gauge Lax b}
\AA(\lambda_-) = \partial_{\sigma} g  g^{-1} + g \L(\lambda_-) g^{-1}.
\end{equation}
\end{subequations}
Since the fields $j_0$ and $j_1$ both take values in $\f$ we have $j_a^{\dag} = - j_a$ for $a = 0, 1$, from which the reality condition on the Lax matrix follows
\begin{equation} \label{real Lax}
\L(\lambda)^{\dag} = - \L(\overline{\lambda}).
\end{equation}
In particular this means that $\L(\lambda_+)^{\dag} =  - \L( \lambda_-)$ which combined with \eqref{gauge Lax} yields
\beq
\AA(\lambda_-) = - \AA(\lambda_+)^\dagger. \label{june2b}
\eeq
This implies, firstly, that $\AA(\lambda_-)$ belongs to the lower Borel subalgebra $\b^-$ of $\f^{\CC}$, or more precisely to $\h_0 \oplus \n^-$. Secondly, since the restriction of $\AA(\lambda_+)$ to the Cartan subalgebra $\h$ of $\f^{\CC}$ is actually contained in $\h_0$, we have $\AA(\lambda_-) \big|_{\h} = -\AA(\lambda_+) \big|_{\h} $.

Therefore, by a single gauge transformation with parameter $g$ we can ensure that the gauge transformed Lax matrix defined as $\AA(\lambda) = \partial_{\sigma} g  g^{-1} + g \L(\lambda) g^{-1}$ has the property that
\begin{equation} \label{A conditions}
\begin{split}
(i) \quad &\AA(\lambda_{\pm}) \in \b^{\pm},\\
(ii) \quad &\AA(\lambda_-) \big|_{\h} = - \AA(\lambda_+) \big|_{\h}.
\end{split}
\end{equation}

To see why this definition of $g$ is a deformation of the principal chiral field, consider the limit when $\epsilon \to 0$. In this limit, the pair of points $\lambda_{\pm}$ in \eqref{twist poles PCM} degenerate to a single point at $\lambda = 0$. Property $(i)$ then requires that $\AA(0)$ be in both $\b^+$ and $\b^-$ and hence $\AA(0) \in \h$. But then property $(ii)$ implies that $\AA(0) = 0$, which is exactly the defining property of the principal chiral field.

\paragraph{Singularity at $\epsilon = 1$.} When the deformation parameter lies in the range $0 < \epsilon < 1$, the points $\lambda_{\pm}$ defined in \eqref{twist poles PCM} are distinct and the above procedure can be used to define the field $g$. As explained above, when $\epsilon = 0$ the pair of points $\lambda_{\pm}$ merge at $\lambda = 0$ and $g$ becomes identified with the principal chiral field. Likewise, as $\epsilon \to 1$ the pair of points $\lambda_{\pm}$ both move off towards infinity. However, the difference here is that in the limit $\lambda \to \infty$ the Lax matrix vanishes identically and the above procedure for defining $g$ no longer makes sense. As we shall see later, this is a symptom of the fact that the deformed theory is only defined for $0 \leq \epsilon < 1$.

\paragraph{Definition of the conjugate momentum.}

So far we have defined a field $g$ for any value of the deformation parameter $\epsilon$ in the range $0 \leq \epsilon < 1$, which identifies in the limit $\epsilon \to 0$ with the principal chiral field. In order to describe the dynamics of this new field $g$ we shall need to relate it to the components $(j_0, j_1)$ of the current whose dynamics is known, and in fact independent of $\epsilon$. In analogy with the Hamiltonian analysis of the principal chiral model, this requires introducing another field $X$ which will essentially turn out to be the conjugate momentum of $g$. We will then be able to express $(j_0, j_1)$ in terms of the pair of Hamiltonian fields $(g, X)$.

We therefore define
\begin{equation} \label{field X}
X = \frac{i}{2 \gamma} \big( \AA(\lambda_+) - \AA(\lambda_-) \big),
\end{equation}
where the parameter $\gamma$ is a normalisation to be fixed later. In the limit $\epsilon \to 0$, this expression has to identify with the derivative of $\AA(\lambda)$ in $\lambda$ evaluated at $\lambda=0$. In view of \eqref{twist poles PCM} this fixes the leading behaviour of $\gamma$ to be $\gamma \sim -\epsilon$ as $\epsilon \to 0$. Furthermore, due to the property \eqref{june2b}, we have $X^{\dag} = -X$ and therefore $X$ takes values in $\f$ provided $\gamma$ is real.

\subsection{The deformed model}
\label{secpcm24}
\paragraph{Non-split $R$-matrix.} Equation \eqref{field X} expresses $X$ as a difference of the quantities $\AA(\lambda_{\pm})$ taking values in the Borel subalgebras $\b^{\pm}$ of $\f^{\CC}$. It turns out to be possible to invert this relation so as to express both $\AA(\lambda_{\pm})$ in terms of $X$ by introducing a certain $\mathbb{R}$-linear operator on $\f$.

To define this operator we begin by expressing the quantities $\AA(\lambda_{\pm})$ satisfying the properties \eqref{A conditions} in terms of basis elements, namely
\begin{equation} \label{A components}
\AA(\lambda_{\pm}) = \pm \gamma \bigg( \sum_{i = 1}^n h_i H^i + \sum_{\alpha > 0} e_{\pm \alpha} E^{\pm \alpha} \bigg).
\end{equation}
We may then write $X$ as defined by \eqref{field X} more explicitly in terms of the basis \eqref{compact generators} of $\f$ as
\begin{equation}
X = \sum_{i = 1}^n h_i T^i + \frac{1}{2 \sqrt{2}} \sum_{\alpha > 0}
\Bigl( (e_{\alpha} + e_{-\alpha}) B^{\alpha} + i (e_{\alpha} - e_{-\alpha}) C^{\alpha}
 \Bigr). \label{june6c}
\end{equation}
Using the reality condition \eqref{herm conj} we find $e_{- \alpha} = \overline{e}_{\alpha}$ so that all the above components of $X$ in this basis are indeed real. If we now introduce an $\mathbb{R}$-linear operator $R : \f \to \f$ as follows \cite{Klimcik:2008eq}
\begin{equation} \label{R def}
R(T^i) = 0, \qquad R(B^{\alpha}) = C^{\alpha}, \qquad R(C^{\alpha}) = - B^{\alpha},
\end{equation}
then the sum of the quantities $\AA(\lambda_{\pm})$ is given simply by
\begin{equation} \label{field RX}
R X = \frac{1}{2 \gamma} \big( \AA(\lambda_+) + \AA(\lambda_-) \big).
\end{equation}
The $\mathbb{R}$-linear map defined in \eqref{R def} is an $R$-matrix of the so called `non-split' type since it satisfies the following variant of the modified classical Yang-Baxter equation
\begin{equation} \label{mCYBE non-split}
[RM, RN] - R\big( [RM, N] + [M, RN] \big) = [M,N].
\end{equation}
We refer to appendix \ref{app: mCYBE} for a brief comparison of the properties of the $R$-matrix introduced here with the $R$-matrix of the `split' type used, for instance, in \cite{Delduc:2012qb}.
 Finally, combining equations \eqref{field X} and \eqref{field RX} we may solve the pair of conditions \eqref{A conditions} and write
\begin{equation} \label{A from R X}
\AA(\lambda_{\pm}) = \gamma (R \mp i) X.
\end{equation}

\paragraph{Lifting to $(g, X)$.} It is now possible to explicitly relate the fields $(g, X)$ introduced previously to the fields $(j_0, j_1)$ used thus far. Doing so will, in particular, enable us to describe the Hamiltonian dynamics of $(g,X)$.  Substituting the relation \eqref{A from R X} into the expressions \eqref{gauge Lax} for the gauge transformed Lax matrix at the points $\lambda_{\pm}$ we obtain
\begin{equation} \label{Lax KAN}
\L(\lambda_{\pm}) = - g^{-1} \partial_{\sigma} g + \gamma \, g^{-1} \big( (R \mp i) X \big) g.
\end{equation}
On the other hand, the Lax matrix at these points can certainly be obtained directly in terms of the fields $(j_0, j_1)$ since
\begin{equation*}
\L(\lambda_{\pm}) = \frac{1}{1 - \lambda_{\pm}^2} (j_1 + \lambda_{\pm} \, j_0) = (1 - \epsilon^2) \, j_1 \pm i \epsilon \sqrt{1 - \epsilon^2} \, j_0.
\end{equation*}
Comparing the above two expressions for $\L(\lambda_{\pm})$ immediately yields the desired expressions for $(j_0, j_1)$ in terms of $(g, X)$, namely
\begin{align*}
j_1 &= \frac{1}{1 - \epsilon^2} \big( - g^{-1} \partial_{\sigma} g + \gamma \, g^{-1} (R X) g \big),\\
j_0 &= - \frac{\gamma}{\epsilon \sqrt{1 - \epsilon^2}} \, g^{-1} X g.
\end{align*}
If we fix
$\gamma = -\epsilon (1 - \epsilon^2)^{3/2}$ then one can show that the full list of deformed Poisson brackets \eqref{interpolating PB PCM} for the components of the current $(j_0, j_1)$ follows from the above relations and the following Poisson brackets for $g$ and $X$,
\begin{subequations} \label{gX PB}
\begin{align}
\{g_{\1}(\sigma), g_{\2}(\sigma') \}_{\epsilon} &=0,\\
\label{gX PB a} \{ X_{\1}(\sigma), X_{\2}(\sigma') \}_{\epsilon} &= \big[ C_{\1\2}, X_{\2}(\sigma) \big] \delta_{\sigma \sigma'},\\
\label{gX PB b} \{ X_{\1}(\sigma), g_{\2}(\sigma') \}_{\epsilon} &= C_{\1\2} \, g_{\2}(\sigma) \delta_{\sigma \sigma'}.
\end{align}
\end{subequations}
To establish this result, one needs to use the fact that $R$ is a non-split anti-symmetric $R$-matrix. This enables in particular to derive the following useful intermediate results,
\begin{subequations}
\begin{align*}
\{ (g^{-1}RXg)_{\1}(\sigma), (g^{-1}RXg)_{\2}(\sigma')\}_{\epsilon} &= [C_{\1\2}, (g^{-1}Xg)_{\2}]\dss,\\
\{  (g^{-1}\partial_\sigma g)_{\1}(\sigma), (g^{-1}RXg)_{\2}(\sigma')
 \}_{\epsilon} &=   -\{ (g^{-1}RXg)_{\1}(\sigma), (g^{-1}\partial_{\sigma'} g)_{\2}(\sigma')  \}_{\epsilon}.
\end{align*}
\end{subequations}
The final expressions for the components $(j_0, j_1)$ in the deformed theory read
\begin{subequations} \label{j from g X PCM}
\begin{align}
j_1 &= - \frac{1}{1 - \epsilon^2} g^{-1} \partial_{\sigma} g - \epsilon \sqrt{1 - \epsilon^2} \, g^{-1} (R X) g,\\
j_0 &= (1 - \epsilon^2) \, g^{-1} X g.
\end{align}
\end{subequations}
We clearly see from these expressions that when $\epsilon \to 0$ we obtain the relation $j_1 = - g^{-1} \partial_{\sigma} g$ of the principal chiral model. On the other hand we also obtain
  $j_0 =  g^{-1} X g$ which identifies $X$ with the  component $l_0 = g  j_0 g^{-1}$ of the right invariant current in this limit. In particular, we see that the Poisson algebra \eqref{gl0 PB} remains undeformed when $\epsilon \neq 0$ since \eqref{gX PB} is exactly of the same form. Note by contrast that we no longer have $\{ j_1(\sigma), g(\sigma') \}_\epsilon =0$ when $\epsilon \neq 0$.

As previously anticipated, we explicitly observe the presence in \eqref{j from g X PCM} of a singularity at $\epsilon = 1$. In particular, if we insert the relations \eqref{j from g X PCM} into the Hamiltonian $H_{\rm PCM}$ of the principal chiral model we find that the resulting Hamiltonian of the deformed model is singular at $\epsilon = 1$.

Finally, the equations of motion for $g$ and $X$ are obtained
by computing their Poisson brackets with the Hamiltonian $H^{\epsilon}$.
One finds
\begin{subequations}
\begin{align}
\partial_{\tau} g g^{-1} &= - (1-\epsilon^2)^2 \Bigl( 1 - \frac{\epsilon^2}{1-\epsilon^2}R^2\Bigr) X
+ \frac{\epsilon}{\sqrt{1-\epsilon^2}} R(\partial_\sigma g g^{-1}), \label{june3a} \\
\partial_{\tau} X &=   \frac{1}{1-\epsilon^2} \partial_\sigma(g j_1 g^{-1})
- \epsilon \sqrt{1-\epsilon^2} \big( [ R(gj_1g^{-1}), X]
+ [ gj_1g^{-1}, RX] \big). \label{july7a}
 \end{align}
\end{subequations}

\subsection{Symmetry algebra} \label{sec: symmetry pcm}

Having completely defined the deformed model in the Hamiltonian formalism, we now turn to the description of its symmetries. In the principal chiral model,
the generators of the global $F_L \times F_R$ symmetry can be conveniently
extracted from the leading expansion of the monodromy at $\lambda = 0$
and $\lambda = \infty$, respectively. We will show that the symmetries of the
deformed model with $\epsilon \neq 0$ can be similarly obtained by expanding the monodromy
but at the points $\lambda = \lambda_{\pm}$ and $\lambda = \infty$.

\paragraph{Undeformed $F_R$ symmetry.}

To begin with, consider the expansion of the Lax matrix at the point $\lambda =
\infty$. At leading
order it is given simply by \eqref{Lax at 0}, namely
\begin{equation*}
\L(\lambda) = - \lambda^{-1} j_0 + O(\lambda^{-2}).
\end{equation*}
Thus the expansion of the monodromy at $\lambda = \infty$ will start with the same
local charges
$\int d\sigma j_0$ as in the undeformed theory. However, referring back to the
deformed Poisson algebra \eqref{interpolating PB a} we see that it is natural to scale
these charges for $\epsilon \neq 0$ by defining
\begin{equation} \label{QR def}
Q^R = \frac{1}{1 - \epsilon^2} \int d\sigma j_0.
\end{equation}
The charges \eqref{QR def} so defined then satisfy the same Poisson algebra at all
values of the deformation parameter $\epsilon$. Moreover, these charges generate the same $F_R$ symmetry on the group element $g$.

\paragraph{Deformed $F_L$ symmetry: Charges.}

Next, we consider how the $F_L$ symmetry of the principal chiral model is affected by the deformation. We shall do this in two steps. We start by identifying the relevant conserved
charges and subsequently proceed to determine their Poisson algebra.

A convenient way of extracting these charges in the principal chiral model is to
first perform a gauge transformation by the principal chiral field and then read off
the charges from the expansion of the gauge transformed monodromy at $\lambda =
0$. As explained above, the double pole of the twist function
at $\lambda = 0$ gets replaced in the deformed theory by the two single poles at $\lambda = \lambda_{\pm}$ of the deformed twist function. In light of all this, a natural
prescription for extracting the
corresponding charges in the deformed theory is to first perform
a gauge transformation by the group valued field $g$ and consider the expansion of the gauge transformed monodromy at the points $\lambda = \lambda_\pm$.

We shall therefore consider the expansions of the
gauge transformed Lax matrix $\AA(\lambda)$ around $\lambda_\pm$. The first thing
to note is that since
the leading terms of these expansions are non-zero, the extraction of the
corresponding charges is far more involved. This is to be contrasted with the
situation in the principal chiral model where the expansion of the gauge
transformed Lax matrix at $\lambda = 0$ starts with  $\AA(\lambda) =- \lambda
l_0 + O(\lambda^2)$.
However, the important point is that although $\AA(\lambda_\pm)$ are both non-zero,
they each live in a Borel subalgebra of $\f^{\CC}$. This will enable us to extract
individual charges directly from the path ordered exponential entering the definition
of the gauge transformed monodromy at these points.

Specifically, if $T(\lambda)$ is the monodromy, then the
gauge transformed monodromy at $\lambda_\pm$ reads
\begin{equation*}
T^g(\lambda_\pm)= g(\infty) T(\lambda_{\pm}) g(-\infty)^{-1} = P \overleftarrow{\exp} \left[
\int_{- \infty}^{\infty} d\sigma \AA(\lambda_{\pm}) \right].
\end{equation*}
Recalling the expressions \eqref{A components} for the gauge transformed Lax matrix, namely
\beqz
\AA(\lambda_{\pm}) = \pm \gamma \bigg( \sum_{i = 1}^n h_i H^i +
\sum_{\alpha > 0} e_{\pm \alpha} E^{\pm \alpha} \bigg),
\eeqz
we will show that the  Cartan components of $\AA(\lambda_\pm)$ can be
factored out of the above path ordered exponential. For this, we
will use the following identity, valid for any
functions $\phi_i$ and $L_{\pm \alpha}$ of $\sigma$,
\begin{multline}
P \overleftarrow{\exp} \biggl[ \int_{\sigma_1}^{\sigma_2} d\sigma \bigg(
 \sum_{i = 1}^n (\partial_{\sigma} \phi_i) H^i + \sum_{\alpha > 0} L_{\pm \alpha}
E^{\pm \alpha} \bigg) \biggr]
= \exp \biggl( \sum_{i = 1}^n \phi_i(\sigma_2) H^i \biggr) \\
\times P\overleftarrow{\exp}
\biggl[ \int_{\sigma_1}^{\sigma_2} d\sigma \sum_{\alpha > 0}
e^{\mp \sum_{i = 1}^n \alpha (H^i) \phi_i(\sigma)}
L_{\pm \alpha}  E^{\pm \alpha} \biggr]
\exp \biggl( - \sum_{i = 1}^n \phi_i(\sigma_1) H^i \biggr). \label{june6b}
\end{multline}
To apply this identity to the path ordered exponential of $\AA(\lambda_+)$ we let
$\phi_i(\sigma) = \int_{-\infty}^\sigma d\sigma' \gamma h_i(\sigma')$
and $L_{\alpha}(\sigma) = \gamma e_{\alpha}(\sigma)$. Then taking $\sigma_1 = - \infty$ and $\sigma_2 = \infty$ in \eqref{june6b} gives
\begin{subequations} \label{A factor Cartan}
\begin{equation} \label{A factor Cartan a}
T^g(\lambda_+) = \exp \biggl( \gamma \int_{- \infty}^{\infty} d\sigma \sum_{i = 1}^n h_i(\sigma)
H^i \biggr)
P \overleftarrow{\exp} \biggl[ \gamma \,
\sum_{\alpha > 0} \int_{- \infty}^{\infty}
d\sigma \, \QQ^E_{\alpha}(\sigma) E^{\alpha} \biggr], \qquad\quad
\end{equation}
where the quantity $\QQ^E_{\alpha}(\sigma)$ is defined below.
Similarly, to describe the path ordered exponential of $\AA(\lambda_-)$ we choose
$\phi_i(\sigma) = \int_{\sigma}^{\infty} d\sigma' \gamma h_i(\sigma')$
and $L_{- \alpha}(\sigma) = - \gamma e_{- \alpha}(\sigma)$. Letting $\sigma_1 = - \infty$ and $\sigma_2 = \infty$ in \eqref{june6b} we obtain
\begin{equation} \label{A factor Cartan b}
T^g(\lambda_-) = P \overleftarrow{\exp} \biggl[ - \gamma \,
\sum_{\alpha > 0} \int_{- \infty}^{\infty}
d\sigma \, \QQ^E_{- \alpha}(\sigma) E^{- \alpha} \biggr]
\exp \biggl( - \gamma \int_{- \infty}^{\infty} d\sigma \sum_{i = 1}^n h_i(\sigma)
H^i \biggr),
\end{equation}
\end{subequations}
where the notation is as follows. For any positive root $\alpha > 0$ we let
\begin{equation}
\QQ^H_{\alpha}(\sigma) = \sum_{i=1}^n \alpha(H^i) h_i(\sigma), \qquad
\QQ^E_{\pm \alpha}(\sigma) = e_{\pm \alpha}(\sigma) \, e^{- \gamma
 \chi_{\alpha}(\sigma)} e^{\gamma \chi_{\alpha}(\mp \infty)}. \label{july7d}
\end{equation}
The function $\chi_{\alpha}$ has the property that $\partial_{\sigma}
 \chi_{\alpha}(\sigma) = \QQ^H_{\alpha}(\sigma)$ and is defined explicitly by
\begin{equation*}
\chi_{\alpha}(\sigma) = \ha \int_{-\infty}^{\infty} d\sigma' \epsilon_{\sigma \sigma'}
\QQ^H_{\alpha}(\sigma') = \int_{-\infty}^{\sigma} d\sigma'
\QQ^H_{\alpha}(\sigma') - \ha \int_{-\infty}^{\infty}
d\sigma' \QQ^H_{\alpha}(\sigma').
\end{equation*}
Here we use the notation $\epsilon_{\sigma \sigma'} = \epsilon(\sigma - \sigma')$
which satisfies $\partial_{\sigma} \epsilon_{\sigma \sigma'} = 2 \delta_{\sigma \sigma'}$.
The boundary values of the function $\chi_{\alpha}$ at $\pm \infty$ are
\begin{equation}
\chi_{\alpha}(\pm \infty) = \pm \ha \int_{-\infty}^{\infty} d\sigma'
\QQ^H_{\alpha}(\sigma'). \label{june6e}
\end{equation}
Note that the transformation $h_i(\sigma) \to \QQ^H_{\alpha_i}(\sigma)$
is invertible since the symmetrized Cartan matrix $B_{ij}$ is invertible, namely we can write
$h_i(\sigma) = \sum_{j = 1}^n B^{-1}_{ij} \QQ^H_{\alpha_j}(\sigma)$ (see appendix \ref{app: real form} for notations).

The advantage of the factorized form \eqref{A factor Cartan} is that the
argument in the remaining path ordered exponential on the right hand side
is nilpotent. Therefore, this path ordered exponential can now be evaluated
explicitly in terms of exponentials of ordinary integrals. In particular,
this allows one to define charges $Q^E_{\alpha}$ corresponding to
each root $\alpha \in \Phi$,
the conservation of which then follows from the conservation of $T^g(\lambda_\pm)$.
Details of the procedure for defining these charges can be found in appendix \ref{app: q-Serre}.
In the remainder of this section, however, we will only be needing the conserved
charges associated with the Cartan generators and the simple roots. These are given by
\begin{equation*}
\int_{-\infty}^{\infty} d\sigma \QQ^H_{\alpha_i}(\sigma) \qquad
\mbox{and} \qquad
\int_{-\infty}^{\infty} d\sigma \QQ^E_{\pm \alpha_i}(\sigma)
\end{equation*}
where the $\alpha_i$, $i = 1, \ldots, n$ are the simple roots of $\f^{\CC}$.

Let us remark that the conservation of the quantities $\int_{-\infty}^{\infty} d\sigma h_i(\sigma)$ could also be shown relatively straightforwardly from their definitions. Indeed,
one can check that the projection of both sides of the equation of motion \eqref{july7a} for $X$ onto $\h$ together with \eqref{june6c} and \eqref{R def}
lead to the desired conservation property.

\paragraph{Deformed $F_L$ symmetry: Algebra.}

In the remainder of this section we determine the Poisson algebra of the charges identified above.

The Lax matrix $\AA(\lambda_{\pm})$ as given in \eqref{A from R X} only depends on the
field $X$, whose expression \eqref{june6c} can be rewritten as
\beq
X =  \sum_{j= 1}^n  i h_j H^j + \frac{i}{2} \sum_{\alpha > 0} (e_\alpha E^{+\alpha}
+ e_{-\alpha} E^{-\alpha}). \label{june6d}
\eeq
It is apparent from this that the Poisson bracket relations of the corresponding charges
will follow solely from the Kostant-Kirillov Poisson bracket
\eqref{gX PB a}.
Using equation \eqref{Casimir HE}, this \pb takes the following form
\beqz
\{ X_{\1} ( \sigma), X_{\2}(\sigma') \}_{\epsilon} =
\Bigl( \sum_{i,j = 1}^n B_{ij}^{-1} H^i \otimes \big[ H^j, X(\sigma) \big] +
\sum_{\alpha > 0} \bigl(E^{\alpha} \otimes \big[ E^{- \alpha}, X(\sigma) \big]  +
E^{- \alpha} \otimes \big[ E^{\alpha}, X(\sigma) \bigr]  \bigr)
\Bigr) \delta_{\sigma \sigma'}.
\eeqz
Comparing coefficients on both sides for the different basis elements of
 $\f^{\CC}$ in the first tensor factor then gives
\begin{equation} \label{hX eX PB}
\{ h_i(\sigma), X(\sigma') \}_{\epsilon} = - i \sum_{j=1}^n B^{-1}_{ij} [ H^j, X(\sigma) ]
\delta_{\sigma \sigma'},\qquad
\{ e_{\pm \alpha}(\sigma), X(\sigma') \}_{\epsilon} = - 2 i [ E^{\mp \alpha}, X(\sigma) ]
\delta_{\sigma \sigma'}.
\end{equation}
Consider the first of these two relations. Using  again \eqref{june6d}
 the comparison of the coefficients of $H^j$ and
$E^{\pm \alpha_j}$ on both sides leads respectively to
\begin{equation*}
\{ h_i(\sigma), h_j(\sigma') \}_{\epsilon} = 0, \qquad
\{ h_i(\sigma), e_{\pm \alpha_j}(\sigma') \}_{\epsilon} = \mp i e_{\pm \alpha_j}(\sigma)
 \delta_{ij} \delta_{\sigma \sigma'}.
\end{equation*}
The second of these relations then implies
\begin{align*}
\{ e^{- \gamma \chi_{\alpha}(\sigma)}, e_{\pm \alpha_j}(\sigma') \}_{\epsilon} &= \pm
\mbox{\small $\frac{i}{2}$} \gamma \, e_{\pm \alpha_j}(\sigma') \alpha(H^j)
e^{- \gamma \chi_{\alpha}(\sigma)} \epsilon_{\sigma \sigma'}.
\end{align*}
Likewise, specialising the second relation in \eqref{hX eX PB} to the simple root $\alpha_i$ and comparing coefficients of $E^{- \alpha_j}$ on both sides gives
\begin{equation*}
\{ e_{\alpha_i}(\sigma), e_{- \alpha_j}(\sigma') \}_{\epsilon} = - 4 i \, \partial_{\sigma}
\chi_{\alpha_i}(\sigma) \delta_{ij} \delta_{\sigma \sigma'}.
\end{equation*}
This allows us to compute commutation relations between the charge densities
$\QQ^E_{\pm \alpha_i}(\sigma)$ and $\QQ^H_{\alpha_i}(\sigma)$, yielding
\begin{subequations} \label{QEH densities alg}
\begin{align}
\big\{ \QQ^E_{\alpha_i}(\sigma), \QQ^E_{- \alpha_j}(\sigma') \big\}_{\epsilon}
&= - 4 i \, \partial_{\sigma} \chi_{\alpha_i}(\sigma) e^{- 2 \gamma \chi_{\alpha_i}(\sigma)}
 \delta_{ij} \delta_{\sigma \sigma'} = 2 i \, \gamma^{-1} \partial_{\sigma} \left(
e^{- 2 \gamma \chi_{\alpha_i}(\sigma)} \right) \delta_{ij} \delta_{\sigma \sigma'},\\
\{ \QQ^H_{\alpha_i}(\sigma), \QQ^E_{\pm \alpha_j}(\sigma') \}_{\epsilon}
&= \mp i B_{ij} \QQ^E_{\pm \alpha_j}(\sigma') \delta_{\sigma \sigma'}.
\end{align}
\end{subequations}
We now define the integrated charges from the above densities, namely
\begin{equation} \label{simple charges}
Q^H_{\alpha_i} = d_i^{-1} \int_{-\infty}^{\infty} d\sigma \QQ^H_{\alpha_i}(\sigma), \qquad
Q^E_{\pm \alpha_i} = D_i \int_{-\infty}^{\infty} d\sigma \QQ^E_{\pm \alpha_i}(\sigma),
\end{equation}
where we define the notational shorthand
\begin{equation} \label{Di}
D_i = \left( \frac{\gamma}{4 \sinh (d_i \gamma)} \right)^{\frac{1}{2}}.
\end{equation}
These normalisations in the charges $Q^E_{\pm \alpha_i}$ have been introduced
for convenience (\emph{c.f.} \cite{Kawaguchi:2011pf}). The Poisson brackets \eqref{QEH densities alg} for the densities then lead to
\begin{subequations} \label{QEH alg}
\begin{align}
\label{QEH alg a} i \{ Q^H_{\alpha_i}, Q^H_{\alpha_j} \}_{\epsilon}
&=0,\\
\label{QEH alg b} i \big\{ Q^E_{+ \alpha_i}, Q^E_{- \alpha_j} \big\}_{\epsilon}
&= \delta_{ij} \frac{q^{d_i Q^H_{\alpha_i}} - q^{- d_i Q^H_{\alpha_i}}}{q^{d_i} - q^{-d_i}},\\
\label{QEH alg c} i \{ Q^H_{\alpha_i}, Q^E_{\pm \alpha_j} \}_{\epsilon}
&= \pm A_{ij} Q^E_{\pm \alpha_j}.
\end{align}
\end{subequations}
Here we have made use of the values \eqref{june6e} and  introduced the new parameter
\begin{equation*}
q = e^{\gamma} = \exp \left( - \epsilon (1 - \epsilon^2)^{\frac{3}{2}} \right).
\end{equation*}
Furthermore, the charges $Q^E_{\alpha_i}$ also satisfy certain $q$-Poisson-Serre relations. To write these down we introduce a $q$-analogue of the deformed Poisson bracket as follows. We say that $Q^E_{\alpha_i}$ defined in \eqref{simple charges} is associated with the simple root $\alpha_i$. Let $A_{\alpha}$ and $A_{\beta}$ denote charges associated with any pair of positive roots $\alpha, \beta > 0$ and define their $q$-Poisson bracket as
\begin{equation} \label{q-bracket}
\big( \text{ad}_{\{ \cdot, \cdot \}_{q \, \epsilon}} A_{\alpha} \big) (A_{\beta} ) \coloneqq \{ A_{\alpha}, A_{\beta} \}_{q \, \epsilon} \coloneqq \{ A_{\alpha}, A_{\beta} \}_{\epsilon} + i \gamma \, (\alpha, \beta) A_{\alpha} A_{\beta}.
\end{equation}
If $\alpha + \beta$ is a root then we regard the resulting quantity $\{ A_{\alpha}, A_{\beta} \}_{q \, \epsilon}$ as being associated with this root. The operator $\big( \text{ad}_{\{ \cdot, \cdot \}_{q \, \epsilon}} A_{\alpha} \big)^n$ may then be defined recursively for any $n \geq 1$. Using this notation, the $q$-Poisson-Serre relations can be written succinctly as follows
\begin{equation} \label{q-Serre}
\big( \text{ad}_{\{ \cdot, \cdot \}_{q \, \epsilon}} Q_{\alpha_i}^E \big)^{1 - A_{ij}} (Q_{\alpha_j}^E) = 0.
\end{equation}
This identity is proved for all classical Lie algebras $\f$ in appendix \ref{app: q-Serre}.

Finally, the charges \eqref{simple charges} have the following behaviour under complex conjugation
\begin{equation} \label{reality Q}
\overline{Q}^H_{\alpha_i} = Q^H_{\alpha_i}, \qquad \overline{Q}^E_{\alpha_i} = q^{- d_i Q^H_{\alpha_i}} Q^E_{- \alpha_i},
\end{equation}
which is easily seen to preserve the relations \eqref{QEH alg}. The $q$-Poisson-Serre relations \eqref{q-Serre} are also mapped to the corresponding relations for negative roots. These take the form
\begin{equation*}
\big( \text{ad}_{\{ \cdot, \cdot \}_{q^{-1} \, \epsilon}} Q_{- \alpha_i}^E \big)^{1 - A_{ij}} (Q_{- \alpha_j}^E) = 0,
\end{equation*}
where similarly to \eqref{q-bracket} we define the $q$-Poisson bracket of any two charges $A_{-\alpha}$ and $A_{- \beta}$ associated with the negative roots $- \alpha, - \beta < 0$ as
\begin{equation*}
\big( \text{ad}_{\{ \cdot, \cdot \}_{q^{-1} \, \epsilon}} A_{- \alpha} \big) (A_{- \beta} ) \coloneqq \{ A_{- \alpha}, A_{- \beta} \}_{q^{-1} \, \epsilon} \coloneqq \{ A_{- \alpha}, A_{- \beta} \}_{\epsilon} - i \gamma \, (\alpha, \beta) A_{- \alpha} A_{- \beta}.
\end{equation*}

\paragraph{Interpretation as semiclassical limit of $U_{\widehat{q}}(\f)$.} The algebra of the deformed $F_L$ symmetry just obtained bears a strikingly resemblance with the relations of the quantum group $U_q(\f)$, but where the commutators are replaced by Poisson brackets. To close the discussion on symmetries, we will show that the Poisson algebra generated by the charges $Q^H_{\alpha_i}$ and $Q^E_{\pm \alpha_i}$, subject to the relations \eqref{QEH alg}, \eqref{q-Serre} and \eqref{reality Q}, coincides exactly with the semiclassical limit $\hbar \to 0$ of the compact real form $U_{\widehat{q}}(\f)$ of the quantum group $U_{\widehat{q}}(\f^{\CC})$ where $\widehat{q} = q^{\hbar}$. The resulting Poisson algebra, which we shall denote $U^{\mathcal{P}}_q(\f)$, gives a one-parameter deformation of the Poisson algebra corresponding to the Lie algebra $\f$. Similar semiclassical limits of finite dimensional quantum groups were considered in \cite{Babelon:1987jg, Babelon:1991ah, Ballesteros_2009} and the case of the quantum affine algebra $U_{\widehat{q}}(\, \widehat{\mathfrak{sl}}_2)$ in \cite{Kawaguchi:2012ve}.

Recall that $U_{\widehat{q}}(\f^{\CC})$ is generated by $\widehat{H}_i, \widehat{E}_i, \widehat{F}_i$ for $i = 1 ,\ldots, n = \text{rk} \, \f^{\CC}$ subject to the relations
\begin{subequations} \label{Uqf rel}
\begin{alignat}{2}
\label{Uqf rel a} [ \widehat{E}_i, \widehat{F}_j ] &= \delta_{ij} \frac{\widehat{K}_i - \widehat{K}_i^{-1}}{\widehat{q}^{d_i} - \widehat{q}^{-d_i}}, &\qquad
[\widehat{H}_i, \widehat{H}_j ] &= 0,\\
\label{Uqf rel b} [ \widehat{H}_i, \widehat{E}_j ] &= A_{ij} \widehat{E}_j, &\qquad
[ \widehat{H}_i, \widehat{F}_j ] &= - A_{ij} \widehat{F}_j,
\end{alignat}
\end{subequations}
where $\widehat{K}_i = \widehat{q}^{d_i \widehat{H}_i}$, together with the $\widehat{q}$-Serre relations which may be written as \cite{Khoroshkin_1991}
\begin{equation} \label{q-Serre Uqf}
\big( \text{ad}_{[ \cdot, \cdot ]_{\widehat{q}}} \widehat{E}_i \big)^{1 - A_{ij}} (\widehat{E}_j) = 0, \qquad
\big( \text{ad}_{[ \cdot, \cdot ]_{\widehat{q}^{-1}}} \widehat{F}_i \big)^{1 - A_{ij}} (\widehat{F}_j) = 0.
\end{equation}
Here we have introduced the $\widehat{q}$-analog of the commutator along with the corresponding $\widehat{q}$-analog of the adjoint action as
\begin{equation} \label{q-commutator}
\big( \text{ad}_{[ \cdot, \cdot ]_{\widehat{q}^{\pm 1}}} \widehat{A}_{\alpha} \big) (\widehat{A}_{\beta} ) \coloneqq [ \widehat{A}_{\alpha}, \widehat{A}_{\beta} ]_{\widehat{q}^{\pm 1}} \coloneqq \widehat{A}_{\alpha} \widehat{A}_{\beta} - \widehat{q}^{\pm (\alpha, \beta)} \widehat{A}_{\beta} \widehat{A}_{\alpha},
\end{equation}
where the $+$ (respectively $-$) sign is used if the roots $\alpha, \beta$ are positive (respectively negative).

There are many possible Hopf algebra structures on $U_q(\f^{\CC})$ corresponding to different choices of coproducts. The real structures on $U_q(\f^{\CC})$ have been classified in \cite{Twietmeyer:1991mj} with respect to the standard coproduct \cite{Chari_Pressley_1994}, but other choices of coproducts lead to alternative reality conditions \cite{Sobczyk_1993}. For our purposes we shall consider the coproduct defined on the generators as \cite{Ruegg:1993eq, Khoroshkin_1991}
\begin{equation} \label{coproduct Uqf}
\Delta \widehat{E}_i = \widehat{E}_i \otimes 1 + \widehat{K}_i^{-1} \otimes \widehat{E}_i, \qquad
\Delta \widehat{F}_i = \widehat{F}_i \otimes \widehat{K}_i + 1 \otimes \widehat{F}_i, \qquad
\Delta \widehat{H}_i = \widehat{H}_i \otimes 1 + 1 \otimes \widehat{H}_i.
\end{equation}
The compact real form $U_{\widehat{q}}(\f)$ of $U_{\widehat{q}}(\f^{\CC})$ then corresponds
to the condition $\widehat{q} \in \mathbb{R}$ and the following choice of
$\ast$-involution on $U_{\widehat{q}}(\f^{\CC})$ \cite{Ruegg:1993eq}
\begin{equation} \label{Uqf compact}
\widehat{K}_i^{\ast} = \widehat{K}_i, \qquad \widehat{E}_i^{\ast} = \widehat{K}^{-1}_i \widehat{F}_i, \qquad \widehat{F}_i^{\ast} = \widehat{E}_i \widehat{K}_i.
\end{equation}

To take the semiclassical limit $\hbar \to 0$ of the above relations we suppose the generators $\widehat{H}_i$, $\widehat{E}_i$ and $\widehat{F}_i$ have the following leading order behaviour in this limit
\begin{equation*}
\hbar \widehat{H}_i \longrightarrow Q^H_{\alpha_i}, \qquad
\hbar \widehat{E}_i \longrightarrow \left( \frac{\sinh(d_i \gamma)}{d_i \gamma} \right)^{\frac{1}{2}} Q^E_{\alpha_i}, \qquad
\hbar \widehat{F}_i \longrightarrow \left( \frac{\sinh(d_i \gamma)}{d_i \gamma} \right)^{\frac{1}{2}} Q^E_{- \alpha_i}.
\end{equation*}
Moreover, we also assume the leading behaviour of the commutator to correspond to the deformed Poisson bracket \eqref{deformed PB PCM}, namely
\begin{equation*}
\frac{1}{\hbar} [ \cdot, \cdot] \longrightarrow i \{ \cdot, \cdot \}_{\epsilon}.
\end{equation*}
It is now easy to see that the relations \eqref{Uqf rel b} reproduce \eqref{QEH alg c} in the limit $\hbar \to 0$. Furthermore, owing to the normalisations of the generators $\widehat{E}_i$ and $\widehat{F}_i$ in this limit we recover also \eqref{QEH alg b} from the first relation in \eqref{Uqf rel a}.

Using the relation $\widehat{q} = q^{\hbar}$, we find that the leading behaviour of the $\widehat{q}$-commutator \eqref{q-commutator} is given by the $q$-Poisson bracket \eqref{q-bracket}, that is
\begin{equation*}
\frac{1}{\hbar} [ \cdot, \cdot]_{\widehat{q}^{\pm 1}} \longrightarrow i \{ \cdot, \cdot \}_{q^{\pm 1} \, \epsilon}.
\end{equation*}
It directly follows from this that the semiclassical limit of the $\widehat{q}$-Serre relations \eqref{q-Serre Uqf} is exactly the $q$-Poisson-Serre relations \eqref{q-Serre}.
The reality conditions \eqref{Uqf compact} also lead to \eqref{reality Q} in this limit.

Finally, taking the semiclassical limit of the coproduct \eqref{coproduct Uqf} we obtain
\begin{subequations} \label{coproduct}
\begin{align}
\Delta Q^E_{\alpha_i} &= Q^E_{\alpha_i} \otimes 1 + q^{- d_i Q^H_{\alpha_i}} \otimes Q^E_{\alpha_i}, \\
\Delta Q^E_{-\alpha_i} &= Q^E_{-\alpha_i} \otimes q^{d_i Q^H_{\alpha_i}} + 1 \otimes Q^E_{-\alpha_i}, \\
\Delta Q^H_{\alpha_i} &= Q^H_{\alpha_i} \otimes 1 + 1 \otimes Q^H_{\alpha_i}.
\end{align}
\end{subequations}
Equipped with this choice of coproduct, the real Poisson algebra $U^{\P}_q(\f)$, defined by the relations \eqref{QEH alg}, \eqref{q-Serre} and the real structure \eqref{reality Q} on the charges $Q^H_{\alpha_i}$ and $Q^E_{\pm \alpha_i}$, acquires the structure of a real Poisson-Hopf algebra.

\subsection{Yang-Baxter $\sigma$-model}
\label{secact}
In this subsection, we will show that the deformed model coincides with the
Yang-Baxter $\sigma$-model introduced by \Klimcik in \cite{Klimcik:2002zj,Klimcik:2008eq}.
For this we need to perform the inverse Legendre transform from the Hamiltonian formalism
to the Lagrangian formalism.

\paragraph{Lagrangian.}
The inverse Legendre transform is given by
\beq
L = \kappa(\partial_{\tau} g g^{-1}, X) - h^\epsilon \label{june3k}
\eeq
where the Hamiltonian density $h^\epsilon = h^{PCM}$, defined by equation \eqref{H PCM}, can be
re-expressed in terms of the light-cone components $j_\pm = j_0 \pm j_1$ of the current as
\beq
h^\epsilon = -\qa \kappa(j_+,j_+) - \qa \kappa(j_-,j_-). \label{june3b}
\eeq
As usual, to rewrite \eqref{june3k} in terms of Lagrangian fields we begin by expressing $X$ in terms of $g$ and its time derivative. This
can be done using the equation of motion \eqref{june3a}.
It turns out to be convenient to express everything in terms of the following variable
\beq
\eta = \frac{\epsilon}{\sqrt{1-\epsilon^2}}. \label{june3l}
\eeq
Noting that $1 \pm \eta R$ is invertible since $R$ is a real skew-symmetric operator and therefore
has only imaginary eigenvalues, one obtains
\beq
X= -\ha (1 + \eta^2)^2 \biggl( \frac{1}{1-\eta R} \partial_- g g^{-1} +
\frac{1}{1+\eta R} \partial_+ g g^{-1}\biggr) \label{june3f}
\eeq
with $\partial_\pm = \partial_{\tau} \pm \partial_\sigma$.
  Using this result we may also express $j_\pm$ in terms of Lagrangian fields.
Starting from equations \eqref{j from g X PCM} we have
\beq
j_\pm = \mp \frac{1}{1-\epsilon^2} g^{-1}\partial_{\sigma} g + (1-\epsilon^2) g^{-1}
(1\mp \eta R) X g. \label{june3d}
\eeq
Then combining equations \eqref{june3d} and \eqref{june3f} we find
\beq
g j_\pm g^{-1} = - \frac{1}{1-\epsilon^2} \frac{1}{1\pm \eta R} \partial_\pm g g^{-1}. \label{june3aa}
\eeq
The last step consists in substituting  \eqref{june3f} and \eqref{june3aa}  into the expression  \eqref{june3k} for the Lagrangian. This  yields the Lagrangian of the deformed model,
\beq
L =-\ha  \kappa \biggl( \partial_+g g^{-1}, \frac{(1 + \eta^2)^2}{1 - \eta R}  \partial_- g g^{-1} \biggr)
\label{june2c}
\eeq
where the operator $R$ is the non-split $R$-matrix defined by equation \eqref{R def}
and $\eta$ is expressed in terms of the deformation parameter $\epsilon$ as \eqref{june3l}. This corresponds to the Yang-Baxter $\sigma$-model defined by
\Klimcik in \cite{Klimcik:2002zj,Klimcik:2008eq}.
Finally, note that when $\epsilon$ tends to zero the Lagrangian \eqref{june2c} reduces
to that of the principal chiral model.

\paragraph{Comments.} To close our discussion on the deformation of the principal chiral model, we compare our definition of the field $g$ given in section \ref{june3ad} with the corresponding definition in \cite{Klimcik:2008eq}.

Consider the extended solution $\Psi(\lambda, \sigma)$ of the principal chiral model, which by definition solves the auxiliary linear problem
\begin{equation*}
\partial_{\sigma} \Psi(\lambda) \Psi(\lambda)^{-1} = \L(\lambda), \qquad \Psi(\lambda, 0) = {\bf 1}.
\end{equation*}
Since the Lax matrix \eqref{Lax PCM} has the property that $\L(0) = j_1 = - g^{-1} \partial_{\sigma} g$, it follows that the principal chiral field $g$ (or rather its inverse) can be recovered from the extended solution evaluated at $\lambda = 0$, namely $g^{-1} = \Psi(0)$. Similarly, it was shown in \cite{Klimcik:2008eq} that the field $g$ of the Yang-Baxter $\sigma$-model can also be retrieved from the same extended solution $\Psi(\lambda)$ of the principal chiral model, but evaluated instead at the special point $\lambda = i \eta$. More precisely, $g^{-1}$ coincides with the element of the compact subgroup $F$ in the Iwasawa decomposition of $\Psi(i \eta) \in F^{\CC}$. To see why this definition agrees with ours, note first that $i \eta$ corresponds to the pole $\lambda_+$. Letting $\Psi(i \eta) = g^{-1} a n$ be the Iwasawa decomposition, where $g$ and $a n$ respectively take values in $F$ and the Borel subgroup $B^+ = \exp \b^+ \subset F^{\CC}$, we may then write
\begin{equation*}
\L(i \eta) = \partial_{\sigma} \Psi(i \eta) \Psi(i \eta)^{-1} = \partial_{\sigma} (g^{-1}) g + g^{-1} \big( \partial_{\sigma}(an) (an)^{-1} \big) g.
\end{equation*}
But this agrees precisely with \eqref{Lax KAN} which can be rewritten as
\begin{equation*}
\L(i \eta) = \partial_{\sigma} (g^{-1}) g + g^{-1} \big( \gamma (R - i) X \big) g.
\end{equation*}
In particular, we have the identification $\partial_{\sigma}(an) (an)^{-1} = \gamma (R - i) X$ as elements in $\b^+$.

It is now apparent that one of the virtues of our approach lies in the identification of the special points $\pm i \eta$ with the poles of the twist function. This will be fully exploited in the next section to extend the above analysis and construct an integrable deformation of coset $\sigma$-models.

\section{Deforming symmetric space $\sigma$-models}
\label{secdeformedcoset}

In this section we discuss the deformation of symmetric space $\sigma$-models,
following a very similar approach to the one developed
in the previous section
for deforming the principal chiral model. For this reason, we insist more
on the new aspects related to the case at hand and omit details which
are similar to the previous case. We use the conventions and notations
of \cite{Delduc:2012qb}.

\subsection{Symmetric space $\sigma$-models}

\paragraph{Hamiltonian and Lax matrix.}
Let $F$ be a compact Lie group with Lie algebra $\f$. We equip $\f$ with a $\mathbb{Z}_2$-automorphism $\sigma$ so that $\sigma^2 = \text{id}$. This induces the usual decomposition $\f = \f^{(0)} \oplus \f^{(1)}$ into the eigenspaces of $\sigma$ where $\f^{(0)} = \g$ is a Lie subalgebra with corresponding Lie group $G = \exp \g$. Let $P_0$ and $P_1$ denote the projection operators onto the respective subspaces $\f^{(0)}$ and $\f^{(1)}$ relative to this decomposition.

We consider the coset $\sigma$-model on the symmetric space $F/G$. It is described by a pair of fields $A$ and $\Pi$ valued in $\f$. The Poisson structure on the graded components $A^{(0)}, A^{(1)}$ and $\Pi^{(0)}, \Pi^{(1)}$ of these fields reads
\begin{subequations} \label{APi PB}
\begin{align}
\{ A^{(i)}_{\1}(\sigma), A^{(j)}_{\2}(\sigma') \} &= 0,\\
\{ A^{(i)}_{\1}(\sigma), \Pi^{(j)}_{\2}(\sigma') \} &= \big[ C^{(ii)}_{\1\2}, A^{(i+j)}_{\2}(\sigma) \big] \delta_{\sigma \sigma'} - C^{(ii)}_{\1\2} \delta_{ij} \delta'_{\sigma \sigma'},\\
\{ \Pi^{(i)}_{\1}(\sigma), \Pi^{(j)}_{\2}(\sigma') \} &= \big[ C^{(ii)}_{\1\2}, \Pi^{(i+j)}_{\2}(\sigma) \big] \delta_{\sigma \sigma'}.
\end{align}
\end{subequations}
Here $C^{(ii)}_{\1\2}$ are the graded components of the Casimir \eqref{Casimir HE} with respect to the automorphism $\sigma$.

The Hamiltonian of the coset $\sigma$-model is
\begin{equation} \label{Ham coset}
H_{\rm coset} = \int_{-\infty}^{\infty}
d\sigma \left[ T_{++} + T_{--} + \kappa( A^{(0)} , \Pi^{(0)}) +
 \kappa (\ell , \Pi^{(0)}) \right]
\end{equation}
where $T_{\pm\pm} = - \qa \kappa(A_{\pm}^{(1)} ,A_\pm^{(1)})$ and $A_\pm^{(1)} = \Pi^{(1)} \mp A^{(1)}$.
The field $\ell$ is a Lagrange multiplier associated with the constraint
$\Pi^{(0)}$ corresponding to the coset gauge invariance.

The integrability of this model is encoded at the Hamiltonian level in the Lax matrix \cite{Vicedo:2009sn}
\begin{equation} \label{coset Lax}
\L(\lambda) = A^{(0)} + \ha (\lambda^{-1} + \lambda) A^{(1)} +
\ha (1 - \lambda^2) \Pi^{(0)} + \ha (\lambda^{-1} - \lambda) \Pi^{(1)}.
\end{equation}
It has the following property with respect to the automorphism
\begin{equation} \label{Lax auto}
\L(- \lambda) = \sigma \big( \L(\lambda) \big),
\end{equation}
which amounts to saying that $\L(\lambda)$ takes values in the twisted loop algebra $\hf^{\sigma}$.

\paragraph{Symmetry algebra.} The coset $\sigma$-model on $F/G$
is invariant under a global $F_L$ symmetry.
The corresponding conserved charges may be extracted
from the leading behaviour of the monodromy matrix at the point
$\lambda = 1$. Indeed, the expansion of the Lax matrix at this point reads
\beq
\L(\lambda) = A - (\lambda - 1) \, \Pi + O\big( (\lambda - 1)^2 \big). \label{june19b}
\eeq
If we introduce the group valued field $g$ through the relation $A = - g^{-1} \partial_{\sigma} g$, and on which the $F_L$ symmetry acts as $g \mapsto U_L g$, then the gauged transformed Lax matrix $\L^g(\lambda)$ previously defined in \eqref{Lax PCM 0} satisfies
\begin{equation}
\L^g(\lambda) = (\lambda - 1) \, X + O\big( (\lambda - 1)^2 \big) \label{june19a}
\end{equation}
with
$X = - g \Pi g^{-1}$.
The coefficient of $\lambda - 1$ in the
expansion of the gauge transformed monodromy matrix then yields
the generator of the $F_L$ symmetry, namely $Q^L = \int d\sigma X$.

\subsection{Setting up the deformation}

\paragraph{Poisson bracket.} Following the same strategy as for the principal
chiral model, we will deform the Poisson bracket $\{ \cdot, \cdot \}$ of $(A, \Pi)$ in \eqref{APi PB} by adding to it the generalized Faddeev-Reshetikhin Poisson
bracket $\{ \cdot, \cdot \}'$ introduced recently in \cite{Delduc:2012qb}. Since these
two brackets are compatible, any linear combination still defines a
Poisson bracket. We therefore set
\begin{equation} \label{deformed PB coset}
\{ \cdot, \cdot \}_{\epsilon} \coloneqq \{ \cdot, \cdot \} + \epsilon^2 \{ \cdot, \cdot \}'
\end{equation}
where $\epsilon$ is a positive real deformation parameter.
The explicit form of this Poisson bracket on the fields $A$ and $\Pi$ is given
in appendix \ref{app: int alg}.

\paragraph{Lax matrix and Hamiltonian.} We shall suppose, as
we did in the \pcm case, that the dependence of the
Lax matrix $\L(\lambda)$ on the fields $(A,\Pi)$ does not change with $\epsilon$.
Moreover, we also impose that the dynamics of the
fields $(A, \Pi)$ remains the same as we turn on
the deformation. These two requirements ensure that the dynamics
of the deformed model remains integrable for $\epsilon \neq 0$.

Therefore, the Hamiltonian $H^{\epsilon}$, which generates the same dynamics on the
fields $(A, \Pi)$ as the coset $\sigma$-model
but with respect to the interpolating bracket $\{ \cdot, \cdot \}_{\epsilon}$,
should satisfy
\begin{equation} \label{ham C eq}
\{ H^{\epsilon}, \L \}_{\epsilon} = \{ H_{\rm coset}, \L \}.
\end{equation}
Postulating a general quadratic ansatz for $H^{\epsilon}$ we find the unique solution of this equation to be
\begin{equation} \label{ham C}
H^{\epsilon} = H_{\rm coset} +
\epsilon^2 \int d\sigma \; \kappa \bigl( \Pi^{(0)}, \Pi^{(0)} \big).
\end{equation}
Plugging \eqref{ham C} directly into \eqref{ham C eq} and using the
fact that $\Pi^{(0)}$ has a vanishing generalised Faddeev-Reshetikhin Poisson
bracket with every function of $(A, \Pi)$, we see that the property \eqref{ham C eq} boils down to the following relation
\beqz
\{ H_{\rm coset}, \L \}' = \Bigl\{ -\int d\sigma \;  \kappa \bigl( \Pi^{(0)}, \Pi^{(0)} \big),
 \L \Bigr\},
\eeqz
which can be checked directly.
Note that the Hamiltonian \eqref{ham C} satisfies the equation \eqref{ham C eq}
strongly. That is to say, the equations of motion generated
by the original Hamiltonian $H_{\rm coset}$ with respect to
the original Poisson bracket $\{ \cdot, \cdot \}$ are reproduced exactly,
including terms proportional to the Hamiltonian constraint $\Pi^{(0)}$.

\subsection{Defining the group valued field}
\label{sec-defofg-coset}

So far we have merely discussed the dynamics of the coset $\sigma$-model
with respect to the deformed Poisson bracket at the level of the fields $(A,\Pi)$.
Following our procedure in the case of the principal chiral model, we anticipate the group
valued field $g$ in the deformed theory to correspond to the parameter of a gauge transformation
of some sort. In fact, it is clear from the discussion leading to equation
\eqref{june19a} that the field $g$ of the coset $\sigma$-model may be described as the
parameter of a gauge transformation sending the Lax matrix $\L(1)$ at $\lambda = 1$ to zero.
To see how such a definition may be deformed when $\epsilon \neq 0$, we turn to the study of the twist function.

\begin{figure}
\centering
\def\svgwidth{100mm}
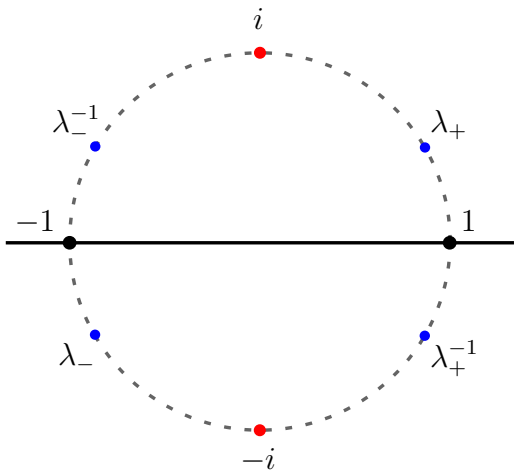
\caption{The four poles $\lambda_{\pm}$, $\lambda_{\pm}^{-1}$ of the twist function $\varphi_{\epsilon}(\lambda)$ for $\epsilon \in [0, 1]$.}
\label{fig: poles coset}
\end{figure}

\paragraph{Deformed twist function and its poles.} As before, the twist function of the
deformed Poisson bracket can be expressed in terms of the twist
functions of the two compatible Poisson brackets,
namely \cite{Delduc:2012qb}
\begin{equation} \label{twists sM gFR}
\varphi_{\sigma}(\lambda) = \frac{4 \lambda}{(1 - \lambda^2)^2}, \qquad
\varphi_{\rm gFR}(\lambda) = \frac{1}{\lambda}.
\end{equation}
The twist of the deformed model is
then defined through the relation
\begin{equation*}
\varphi^{-1}_{\epsilon} = \varphi^{-1}_{\sigma} + \epsilon^2 \varphi^{-1}_{\rm gFR}.
\end{equation*}
Substituting the definitions \eqref{twists sM gFR} into this relation
we find this twist function to be
\begin{equation*}
\varphi_{\epsilon}(\lambda) = \frac{4 \lambda}{\lambda^4 + (4 \epsilon^2 - 2)
\lambda^2 + 1}.
\end{equation*}
As we learned from the \pcm case, the poles of
$\varphi_{\epsilon}$ will play
an important role in defining the field $g$ in the deformed
theory as well as in extracting symmetry generators of the latter.
In the present case we find a bifurcation in the behaviour of these poles at
the special value $\epsilon = 1$. Specifically, for $0 \leq \epsilon < 1$,
if we define an angle $0 \leq \theta < \frac{\pi}{2}$ by letting
\begin{equation*}
\sin \theta = \epsilon
\end{equation*}
then the four poles $\lambda_\pm$ and $\lambda_\pm^{-1}$
of the twist function $\varphi_{\epsilon}$ are located
on the unit circle, with
\begin{equation} \label{twist poles sM}
\lambda_{\pm} \coloneqq \pm e^{i \theta}.
\end{equation}
The two initial double poles of $\varphi_0$ at $\lambda = \pm 1$ (\emph{i.e.}
$\theta = 0$) therefore split into four distinct simple poles of
$\varphi_{\epsilon}$ as we turn on the deformation parameter $\epsilon$
in the range $0 < \epsilon < 1$. But as $\epsilon$ approaches the
value $1$ (\emph{i.e.} $\theta = \frac{\pi}{2}$), the four poles degenerate
once again into two points at $\lambda = \pm i$. The behaviour of these poles
is depicted in Figure \ref{fig: poles coset}. As we increase $\epsilon$ further,
for $\epsilon > 1$ we find that these double poles split once more into
single poles and move off along the imaginary axis.

Since we are interested in deforming away from the coset
$\sigma$-model, we shall focus on the region $0 \leq \epsilon < 1$.
We will discuss briefly what happens at the special value $\epsilon = 1$ in a moment.

\paragraph{Definition of $g$.}  We would now like to generalise
the procedure used in the case of the principal chiral model for
defining the field $g$ at non-zero values of the deformation
parameter $\epsilon \neq 0$.
The novelty here is that the deformed twist function has four simple
poles $\lambda_+^{\pm 1}$ and $\lambda_-^{\pm 1}$ at generic values
of $\epsilon \neq 0$, which degenerate in the limit $\epsilon \to 0$ to the pair
of double poles at $\lambda = \pm 1$, respectively. However, since the field $g$ of
the coset $\sigma$-model is extracted from the point $\lambda = 1$
alone, it is natural to focus only on the points $\lambda_+^{\pm 1}$ for
the purpose of extracting the field $g$ at $\epsilon \neq 0$.

Owing to the reality conditions $A^{\dag} = - A$ and $\Pi^{\dag} = - \Pi$
we have for the Lax matrix
\begin{equation} \label{Lax real coset}
\L(\lambda)^{\dag} = (A^{(0)})^{\dag} + \ha (\overline{\lambda}^{-1} +
\overline{\lambda}) (A^{(1)})^{\dag} + \ha (1 - \overline{\lambda}^2)
(\Pi^{(0)})^{\dag} + \ha (\overline{\lambda}^{-1} - \overline{\lambda})
(\Pi^{(1)})^{\dag} = - \L(\overline{\lambda}).
\end{equation}
In particular, this means that $\L(\lambda_+)^{\dag} = - \L(\lambda_+^{-1})$.
By the exact same reasoning as in section \ref{june3ad} we may argue
here the existence of a field $g \in F$ with the property that the gauge
transformation of the Lax matrix $\L^g(\lambda) \coloneqq
  \partial_{\sigma} g g^{-1} + g \L(\lambda) g^{-1}$ satisfies
\begin{equation} \label{A conditions coset}
\begin{split}
(i) \quad &\L^g(\lambda_+^{\pm 1}) \in \b^{\pm},\\
(ii) \quad &\L^g(\lambda_+) \big|_{\h} = - \L^g(\lambda_+^{-1}) \big|_{\h}.
\end{split}
\end{equation}

The field $g$ so defined has the required property that it reduces
to the field of the coset $\sigma$-model in the limit $\epsilon \to 0$.
Indeed, in this limit the pair of points $\lambda_+^{\pm 1}$ degenerate
to the single point $\lambda = 1$ so that the properties $(i)$ and $(ii)$
together imply that $\L^g(1) = 0$, which is the defining condition of the
coset $\sigma$-model field.

\paragraph{Definition of the conjugate momentum.} Next, we define
a field $X$ taking values in $\f$ which will play the role of the conjugate
momentum of $g$. In exact analogy with the principal chiral case, we
define this field as
\begin{equation} \label{field X coset}
X = \frac{i}{2 \gamma} \big( \L^g(\lambda_+) - \L^g(\lambda_+^{-1}) \big),
\end{equation}
where $\gamma$ is a real normalisation, the dependence of which
on the deformation parameter $\epsilon$ will be fixed later. The reality
condition \eqref{Lax real coset} on the Lax matrix leads to $X^{\dag} = - X$,
therefore ensuring that the field $X$ takes values in $\f$, as desired.

Introducing the same non-split $R$-matrix as in \eqref{R def} we may
then also invert the relation \eqref{field X coset} to express the value
of the Lax matrix at the points $\lambda_+^{\pm 1}$ explicitly as
\begin{equation} \label{A choice coset}
\L^g(\lambda_+^{\pm 1}) = \gamma (R \mp i) X.
\end{equation}

\paragraph{Behaviour at $\epsilon = 1$.}  It turns out that the deformation of the coset
$\sigma$-model that we consider here will only be valid in the range $0 \leq \epsilon < 1$.
To understand what happens at $\epsilon = 1$,
note that the poles of the twist function meet again in pairs $\lambda = \pm i$.
In a neighbourhood of the
point $\lambda = i$ the Lax matrix \eqref{coset Lax} reads
\begin{equation}
\L(\lambda) = A^{(0)} + \Pi^{(0)} - i \Pi^{(1)}-i(\lambda-i)(\Pi^{(0)}+iA^{(1)})+O\big( (\lambda-i)^2 \big).
\label{lax expand i}\end{equation}
Using the deformed Poisson bracket given in Appendix \ref{app: int alg} one easily checks
that the quantities
\begin{equation*}
\hat A=A^{(0)} + \Pi^{(0)} - i \Pi^{(1)},\quad \hat\Pi=\Pi^{(0)}+iA^{(1)}
\end{equation*}
have Poisson brackets at $\epsilon=1$ which are identical to the undeformed Poisson
brackets of the coset $\sigma$-model, namely
\begin{gather*}
\{\hat A_{\1}(\sigma),\hat A_{\2}(\sigma')\}_1=0,\quad
\{\hat \Pi_{\1}(\sigma),\hat\Pi_{\2}(\sigma')\}_1=\big[ C_{\1\2}, \hat\Pi_{\2}(\sigma) \big]
 \delta_{\sigma \sigma'},\\
\{\hat \Pi_{\1}(\sigma),\hat A_{\2}(\sigma')\}_1=
\big[ C_{\1\2}, \hat A_{\2}(\sigma) \big] \delta_{\sigma \sigma'} - C_{\1\2}
\delta'_{\sigma \sigma'}.
\end{gather*}
Notice that equation \eqref{lax expand i} is then completely analogous to equation \eqref{june19b}
which gave the expansion of the Lax matrix around $\lambda=1$.
One can show that the model at $\epsilon =1$ corresponds again to an undeformed coset $\sigma$-model.
However, its fields $(\hat{A}, \hat{\Pi})$ no longer take values in the compact
real form $\f$, but instead satisfy the modified reality condition $\hat
A^\dagger=-\sigma(\hat A)$, $\hat\Pi^\dagger=-\sigma(\hat\Pi)$. In this
case, the group valued field should no longer be taken in the compact Lie group $F$.

\subsection{The deformed model}

In order to describe the dynamics of
the Hamiltonian fields $(g, X)$ we need to relate these to the
fields $(A^{(0,1)}, \Pi^{(0, 1)})$ used up until now.
This is done by expressing the Lax matrix at the points $\lambda_+^{\pm 1} = e^{\pm i \theta}$
in two separate ways.
On the one hand, the definition of the fields $(g, X)$ enable us to write
the gauge transformation of the Lax matrix with parameter $g \in F$
in terms of the field $X \in \f$. Specifically, we have
\begin{equation} \label{coset Lax 1}
\L(e^{\pm i \theta}) = - g^{-1} \partial_{\sigma} g + \gamma \, g^{-1}
\big( (R \mp i) X \big) g.
\end{equation}
On the other hand,
the value of the Lax matrix at these points may also be determined
directly from its definition as
\begin{equation} \label{coset Lax 2}
\L(e^{\pm i \theta}) = A^{(0)} + \cos \theta \, A^{(1)} \mp i e^{\pm i \theta}
\sin \theta \, \Pi^{(0)} \mp i \sin \theta \, \Pi^{(1)}.
\end{equation}
Therefore, equating the two expressions \eqref{coset Lax 1}
and \eqref{coset Lax 2} we find
\begin{equation*}
A^{(0)} + \cos \theta \, A^{(1)} \mp i e^{\pm i \theta} \sin \theta \,
 \Pi^{(0)} \mp i \sin \theta \, \Pi^{(1)} = - g^{-1} \partial_{\sigma} g
+ \gamma \, g^{-1} \big( (R \mp i) X \big) g.
\end{equation*}
Taking the sum and the difference of both sides then yields
\begin{align*}
A^{(0)} + \cos \theta \, A^{(1)} + \sin^2 \theta \, \Pi^{(0)} &= - g^{-1}
\partial_{\sigma} g + \gamma \, g^{-1} (R X) g,\\
\sin \theta \, \Pi^{(1)} + \cos \theta \sin \theta \, \Pi^{(0)} &= \gamma \, g^{-1} X g.
\end{align*}
To extract the individual fields $A^{(0,1)}$ and $\Pi^{(0,1)}$ from
these expressions we should project onto the graded subspaces
$\f^{(0)}$ and $\f^{(1)}$ of the Lie algebra $\f$ using the corresponding
projection operators $P_0$ and $P_1$. This gives
\begin{subequations} \label{API from gX}
\begin{align}
A^{(0)} &= P_0 \big( - g^{-1} \partial_{\sigma} g + \gamma \, g^{-1}
\big( (R - \eta) X \big) g \big),\\
A^{(1)} &= \sqrt{1 + \eta^2} P_1 \big( - g^{-1} \partial_{\sigma} g +
\gamma \, g^{-1} (R X) g \big),\\
\Pi^{(0)} &= \gamma \eta^{-1} (1 + \eta^2) \, P_0 ( g^{-1} X g ), \label{june17a}\\
\Pi^{(1)} &= \gamma \eta^{-1} \sqrt{1 + \eta^2} \, P_1( g^{-1} X g ),
\end{align}
\end{subequations}
where we have defined the variable
\begin{equation*}
\eta = \tan \theta = \frac{\epsilon}{\sqrt{1-\epsilon^2}}.
\end{equation*}
Quite remarkably, one can check that these expressions satisfy the
deformed Poisson algebra  given in appendix \ref{app: int alg}
exactly, if we let
\begin{equation} \label{gamma eq}
\gamma= -\epsilon \, \sqrt{1 - \epsilon^2}=-  \frac{\eta}{1+\eta^2}
\end{equation}
and require the fields $X$ and $g$ to satisfy the exact same Poisson
bracket relations as in the principal chiral model, namely \eqref{gX PB}.

\subsection{Deformed coset $\sigma$-model action}
\label{secggyb}
In this section, we perform the inverse Legendre transform to derive the action
corresponding to our model.

\paragraph{Lagrangian.} The analysis proceeds in exactly the same way as in
subsection \ref{secact}, except for the fact that there is now a constraint. We start with the
definition of the inverse Legendre transform
\begin{equation} \label{Legendre coset}
L = \kappa(\partial_{\tau} g g^{-1}, X) - h^\epsilon=
 \kappa\bigl( (g^{-1} \partial_{\tau} g)^{(1)}, (g^{-1} X g)^{(1)} \bigr) - T_{++}- T_{--}.
\end{equation}
Here we have used equations \eqref{ham C} and \eqref{Ham coset}. Furthermore, we have
imposed the constraint $\Pi^{(0)} \simeq 0$ and made use of its explicit expression \eqref{june17a}.

In order to relate the field $X$ to $(g^{-1} \partial_{\tau} g)^{(1)}$, we compute the time evolution of $g$. First of all, the fields $A^{(1)}_{\pm}$ entering the expression \eqref{Ham coset} for the Hamiltonian can be written in terms of $X$ and $g$ using the relations \eqref{API from gX},
\beq
A_{\pm}^{(1)} = \frac{-1}{\sqrt{1+\eta^2}} P_1\bigl(g^{-1} X g \mp \eta
g^{-1} RX g\bigr) \pm \sqrt{1+\eta^2} (g^{-1} \partial_\sigma g)^{(1)}. \label{june17b}
\eeq
This allows us to express  the Hamiltonian \eqref{ham C} in terms of the fields $g$ and $X$. We may then compute the time evolution of the field $g$ as $g^{-1} \partial_{\tau} g = g^{-1} \{ H^{\epsilon}, g \}_\epsilon$. Extracting the field $X$ from this we find
\begin{equation}
g^{-1} X g \simeq (g^{-1} X g)^{(1)}
=  -\frac{1+\eta^2}{2} \biggl( \frac{1}{1-\eta P_1 \circ R_g} (g^{-1} \partial_- g
)^{(1)} + \frac{1}{1+\eta P_1 \circ R_g} (g^{-1} \partial_+ g)^{(1)}\biggr). \label{june17c}
\end{equation}
Here we have made use once again of the constraint $\Pi^{(0)} \simeq 0$. We have also introduced the operator
\beqz
R_g \coloneqq \text{Ad} \, g^{-1} \circ R \circ \text{Ad} \, g,
\eeqz
which, like $R$ itself, is a non-split solution of the mCYBE \eqref{mCYBE non-split}. Note that $1 \pm \eta P_1 \circ R_g$ is invertible on $\f^{(1)}$ since it is equal to $1 \pm \eta P_1 \circ R_g \circ P_1$ and $P_1 \circ R_g \circ P_1$ is skew-symmetric.

Next, we should also eliminate the field $X$ from $T_{\pm\pm}$ in favour of the Lagrangian
field $g^{-1} \partial_{\tau} g$. For this, we first combine equations
\eqref{june17b} and \eqref{june17c} to get
\begin{equation}
A_{\pm}^{(1)} = \sqrt{1+\eta^2} \frac{1}{1\pm \eta P_1 \circ R_g}(g^{-1} \partial_{\pm} g)^{(1)}.
\label{july7b}
\end{equation}
Then, plugging equations \eqref{june17c} and \eqref{july7b} in  the
inverse Legendre transform \eqref{Legendre coset} yields
\begin{equation} \label{L def cos}
L = -  \ha \kappa\biggl((g^{-1} \partial_+g)^{(1)}, \frac{1+\eta^2}{1-\eta   R_g \circ P_1}
(g^{-1} \partial_- g)^{(1)} \biggr).
\end{equation}
In the limit $\epsilon \to 0$, which corresponds to $\eta \to 0$, one correctly recovers the usual Lagrangian of the $F/G$ coset $\sigma$-model.

\paragraph{Gauge invariance and field equations.} One can check the gauge invariance of the deformed model directly at Lagrangian level. Indeed, under the transformation
\begin{equation} \label{gauge tr}
g(\tau,\sigma) \mapsto g(\tau,\sigma)h(\tau,\sigma), \qquad h(\tau,\sigma)\in G.
\end{equation}
one has the following
\begin{equation*}
(g^{-1} \partial_{\pm} g)^{(1)} \mapsto Ad(h)^{-1}(g^{-1} \partial_{\pm} g)^{(1)},\qquad
R_g \mapsto Ad(h)^{-1}\circ R_g\circ Ad(h).
\end{equation*}
The gauge invariance of the action corresponding to \eqref{L def cos} under \eqref{gauge tr} immediately follows from this.
In particular, for all values of the deformation parameter $\eta$, the physical degrees
of freedom belong to the coset $F/G$. One may also check that the field equations
take the same form as in the coset $\sigma$-model, that is
\begin{equation*}
\left(\partial _+B_-^{(1)} + \bigl[ B_+^{(0)},B_-^{(1)} \bigr] \right)+  \left( \partial _-B_+^{(1)}
+ \bigl[ B_-^{(0)},B_+^{(1)} \bigr] \right)= 0,
\end{equation*}
where the fields $B_{\pm}$ are deformations of $g^{-1}\partial_\pm g$ defined as\footnote{Using the fact that $1 \pm \eta P_1 \circ R_g$ is invertible on $\f^{(1)}$ it follows that $1 \pm \eta R_g \circ P_1$ is invertible on $\f$. Explicitly we have
$\frac{1}{1 \pm \eta R_g \circ P_1} = P_0 + (1 \mp \eta P_0 \circ R_g) \frac{1}{1 \pm \eta P_1 \circ R_g} P_1$.}
\begin{equation} \label{B fields}
B_{\pm} = \frac{1}{1 \pm \eta R_g \circ P_1}(g^{-1}\partial_\pm g).
\end{equation}
Moreover, provided the field equations are satisfied, the fields $A_{\pm} = B^{(0)}_{\pm} + \sqrt{1+\eta^2} B^{(1)}_{\pm}$ satisfy the zero curvature equation
\begin{equation}
\partial_+ A_- - \partial_- A_+ + [A_+, A_-] = 0.  \label{july3a}
\end{equation}
The Lax pair associated with the model just defined may therefore be written as
\begin{equation*}
\L_\pm(\lambda) = A_\pm^{(0)} + \lambda^{\pm 1} A_{\pm}^{(1)}.
\end{equation*}

\subsection{Symmetry algebra}

To end this section we discuss the effect of the deformation on the global $F_L$ symmetry of the coset $\sigma$-model.

Recall that in the case of the principal chiral model, the derivation of the Poisson
algebra \eqref{QEH alg}, \eqref{q-Serre} and the reality conditions \eqref{reality Q}
satisfied by the generators of the deformed $F_L$
symmetry relied solely on the Poisson bracket of the field $X$ with itself in \eqref{gX PB a},
along with the special form \eqref{A from R X} of the Lax matrix at the pair of poles
$\lambda_{\pm}$ of the twist function. The situation in the present case is exactly the
same since the Poisson brackets \eqref{gX PB} are identical and the Lax matrix at the
special points $\lambda_+^{\pm 1}$ takes the similar form \eqref{A choice coset}. The
analysis therefore goes through unchanged in the case at hand, the only difference being
the dependence of the parameter $\gamma$ on $\epsilon$, resulting in a different
expression for $q$. Note also that the corresponding charges
are gauge invariant. This is so because they are built in terms of $X$, which has vanishing
Poisson bracket with $\Pi^{(0)}$.

The deformed coset $\sigma$-model therefore admits a classical $U^{\P}_q(\f)$ symmetry
where the parameter $q$ is now given by
\begin{equation*}
q = e^{\gamma} = \exp \Bigl(- \epsilon \sqrt{1 - \epsilon^2} \Bigr).
\end{equation*}

\section{Deformed $SU(2)/U(1)$ coset $\sigma$-model}
\label{secsu2u1}

As recalled in the introduction, the Lagrangian of the Yang-Baxter $\sigma$-model
\eqref{june2c} on a compact Lie group $F$ reduces in the special case of $F = SU(2)$
to that of the squashed sphere $\sigma$-model. As its name suggests, the target space
of the latter is a certain deformation of the 3-sphere $SU(2) \simeq S^3$. More generally,
however, the deformation is not purely metric since the presence of the $R$-matrix in
the Lagrangian gives rise to a torsion term as well \cite{Klimcik:2008eq}.

For similar reasons, the Lagrangian \eqref{L def cos} of the deformed
coset $\sigma$-model will correspond not only to
a deformation of the metric of the coset $F/G$, but also to the
introduction of torsion in the deformed geometry.  In the present section we consider the
Lagrangian \eqref{L def cos} in the simplest case, which corresponds
to the symmetric space $SU(2)/U(1)$. In this example, since the coset is two dimensional there is no torsion.

\paragraph{Gauge fixed action and equations of motion.}

To begin with we set up some notation. We write the field $g \in SU(2)$ explicitly as
\beqz
g = \left( \!\! \begin{array}{cc}z_1&-\bar z_2\cr z_2& \bar z_1\end{array} \!\! \right),
\qquad \vert z_1\vert^2+\vert z_2\vert^2=1.
\eeqz
Correspondingly, we write a generic element $M$ in the Lie algebra
$\mathfrak{su}(2)$ as
\begin{equation} \label{M def}
M = \left( \! \begin{array}{cc}\alpha&-\bar\beta \cr\beta& -\alpha\end{array} \! \right),\qquad
\bar\alpha=-\alpha.
\end{equation}
The anti-linear anti-involution $M \mapsto M^{\dag}$ is defined here in terms of conjugation and
matrix transposition as $M^{\dag} = \bar{M}^{\top}$. The basis generators
\eqref{compact generators} are proportional to the Pauli matrices, explicitly
$T = i \sigma_3$, $B = i \sigma_1 / \sqrt{2}$ and $C = i \sigma_2 / \sqrt{2}$.
The $\mathbb{Z}_2$-automorphism $\sigma$ of $\mathfrak{su}(2)$ is taken to be
$\sigma(M) = - \sigma_1 M^{\top} \sigma_1$, so that the projectors onto the grade
 $0$ and grade $1$ parts of $M$ are respectively given by
\beqz
P_0 \, M = \left( \! \begin{array}{cc}\alpha&0 \cr 0& -\alpha\end{array} \! \right),\qquad
P_1 \, M = \left( \! \begin{array}{cc}0&-\bar\beta \cr\beta& 0\end{array} \! \right).
\eeqz
Finally, the action of the $R$-matrix defined in \eqref{R def} on the generic element
\eqref{M def} of $\mathfrak{su}(2)$ reads
\beqz
R \, M = \left( \! \begin{array}{cc}0&i\bar\beta \cr i\beta& 0\end{array} \! \right).
\eeqz

To evaluate the Lagrangian \eqref{L def cos} more explicitly we need to invert the
operator $1 - \eta R_g \circ P_1$. A short calculation leads to
\begin{gather*}
\frac{1}{1 - \eta R_g \circ P_1} \, M = \left( \! \begin{array}{cc}\alpha'&
-\bar\beta'\cr\beta'&-\alpha'\end{array} \! \right),\\
\beta' = \frac{\displaystyle \beta}{\displaystyle
1-i\eta(\vert z_1\vert^2-\vert z_2\vert^2)}, \qquad \alpha' = \alpha+i\eta(\beta'\bar
z_1\bar z_2+\bar\beta' z_1z_2).
\end{gather*}
In the case at hand, the model described by the Lagrangian \eqref{L def cos}
is invariant under the right $U(1)$ gauge transformations
\begin{equation*}
g(\sigma, \tau) \mapsto g(\sigma, \tau)
 \left( \!\!
\begin{array}{cc}
e^{i\theta(\sigma, \tau)}&0\cr 0&e^{-i\theta(\sigma, \tau)}
\end{array}
 \!\! \right).
\end{equation*}
We choose to fix this gauge invariance by requiring the component
field $z_1(\sigma, \tau)$ to be real and positive and parameterise
the remaining fields using stereographic coordinates on the sphere.
Hence, we take
\beqz
g =   \frac{1}{\sqrt{1+\bar\psi\psi}} \left( \!\!
 \begin{array}{cc}
 \displaystyle 1 & \displaystyle  -\bar\psi \cr
\displaystyle \psi &1
\end{array}
 \!\! \right).
\eeqz
In terms of the complex field $\psi$, the action associated with the Lagrangian
\eqref{L def cos} then takes the form
\begin{equation}
S[ \psi, \bar\psi ] = \frac{1+\eta^2}{2}\int d\tau d\sigma  \frac{\partial_-
\psi\partial_+\bar\psi+\partial_+\psi\partial_-\bar\psi}
{(1+\bar\psi\psi)^2+\eta^2(1-\bar\psi\psi)^2}.\label{acts2}
\end{equation}
A term which is skew-symmetric in the light-cone coordinates has been omitted
here, since it does not participate in the field equation, which reads
\begin{equation}
\partial_+\partial_-\psi-2\frac{1+\bar\psi\psi-\eta^2(1-\bar\psi\psi)}{(1+\bar\psi\psi)^2
+\eta^2(1-\bar\psi\psi)^2}
\bar\psi\partial_+\psi\partial_- \psi = 0. \label{july3b}
\end{equation}

\paragraph{Zero curvature equation.}

In the chosen gauge, we find that the fields \eqref{B fields} entering the equations of
motion are given by
\begin{gather*}
B_\pm = \left( \! \begin{array}{cc} a_\pm& -\bar b_\pm\cr b_\pm& -a_\pm\end{array} \! \right),\\
b_\pm = \frac{\partial_\pm\psi}{1+\bar\psi\psi\pm i\eta(1-\bar\psi\psi)},\quad a_\pm=
\ha (1\mp i\eta)\bar\psi b_\pm- \ha (1\pm i\eta)\psi\bar b_\pm.
\end{gather*}
In terms of these quantities, the field equation \eqref{july3b}
reduces to the covariant conservation equation
\begin{equation*}
(\partial_+ b_- - 2 a_+ b_-) + (\partial_- b_+ - 2 a_- b_+) = 0.
\end{equation*}
Moreover, provided the field equation is satisfied, one has
\begin{gather*}
(\partial_+ b_- - 2 a_+ b_- ) - ( \partial_- b_+ - 2 a_- b_+) = 0,\\
\partial_+ a_- - \partial_- a_+ - (1 + \eta^2)(\bar b_+ b_- - \bar b_- b_+) = 0,
\end{gather*}
corresponding to the projections on the two gradings of the zero curvature equation
\eqref{july3a}.

\paragraph{Remarks.}

The action \eqref{acts2} has the following interesting property.
It interpolates between the coset $\sigma$-model
on the compact symmetric space $SU(2)/U(1)$ at $\eta=0$ and the coset $\sigma$-model
on the non-compact symmetric space $SU(1,1)/U(1)$ at $\eta=\infty$.
This is reminiscent of the discussion at the end of the subsection
\ref{sec-defofg-coset}. Indeed, the limit $\eta \to \infty$ corresponds
to $\epsilon \to 1$ and we have shown that at this special point, the model
constructed corresponds to an undeformed coset $\sigma$-model.

We end this section by computing, for generic values of $\eta$, the Ricci tensor
associated with the metric $g_{ij}$ appearing in the action \eqref{acts2}. Its only non-vanishing component is given by
\begin{align}
R_{\psi\bar\psi}&=\frac{\partial}{\partial\psi}\frac{\partial}{\partial\bar\psi}
\ln((1+\bar\psi\psi)^2+\eta^2(1-\bar\psi\psi)^2) \notag\\
&= \frac{2(1-\eta^2)}
{(1+\bar\psi\psi)^2+\eta^2(1-\bar\psi\psi)^2}+\frac{16\eta^2\bar\psi\psi}
{((1+\bar\psi\psi)^2+\eta^2(1-\bar\psi\psi)^2)^2}. \label{july3e}
\end{align}
The second term in \eqref{july3e} vanishes in both limits $\eta\rightarrow 0$ and
$\eta\rightarrow\infty$, at which we have $R_{ij}=\pm 4 g_{ij}$ respectively.
It is only in these two limits that one recovers
an Einstein manifold, with opposite curvatures. It is well-known that
the on-shell one-loop divergence in such a model is
proportional to the Ricci tensor \cite{Friedan:1980jf,Friedan:1980jm}. In
the case at hand, such a divergence can be reabsorbed into a renormalization of an
overall factor in front of the action as in the coset $\sigma$-model case and into a renormalization
of the deformation parameter $\eta$.

\section{Conclusion}

In this article we introduced a procedure for constructing integrable deformations of principal chiral models and symmetric space $\sigma$-models associated with compact Lie groups. It is worth emphasising that in this construction, the integrability of the deformed models is obvious from the very outset. Indeed, the deformation originates from the choice of a second Poisson bracket which is compatible with the original one. As such, the generalized Faddeev-Reshetikhin bracket plays an essential role in the initial step of the construction. As in the case of the anisotropic $SU(2)$ principal chiral model, a natural question to consider is whether two-parameter deformations of these $\sigma$-models may also be constructed within this framework using a third compatible Poisson bracket.

Another important ingredient is given by the non-split $R$-matrix which shows up in the resulting Lagrangians. In fact, the integrability of the corresponding field equations relies in a subtle way on the modified classical Yang-Baxter equation for this $R$-matrix.
Its appearance in our construction can be traced back to the fact that the gauge transformed Lax matrix takes values in Borel subalgebras at the poles of the twist function. Moreover, this latter property was essential in order to extract the classical $q$-deformed $U^{\P}_q(\f)$ symmetry algebra.

The charges associated with the $q$-deformed $U_q(\f)$ symmetry were extracted from the leading order behaviour of the monodromy matrix at the poles of the twist function. This raises a natural question with regards to the higher conserved charges. By extracting these from the higher order expansion of the gauge transformed monodromy matrix at the poles of the twist function, we may anticipate that they should satisfy a classical affine $U^{\P}_q(\,\hf\,)$ Poisson-Hopf algebra. Indeed, in the case of the squashed sphere $\sigma$-model, the hidden symmetries were already shown to satisfy a $U^{\P}_q( \, \widehat{\mathfrak{sl}}_2 )$ algebra \cite{Kawaguchi:2012ve}.

Much like the squashed sphere $\sigma$-model, the example of the deformed $SU(2)/U(1)$ coset $\sigma$-model is simple enough that it can be studied very explicitly. In fact, many of the general properties discussed in the general case are also present in this simplest example. This integrable deformation certainly deserves further study.

It is very exciting to consider the possible generalisation of this work. The case of the $AdS_5 \times S^5$ superstring $\sigma$-model, currently under investigation, is particularly enticing, especially because the generalisation of the Faddeev-Reshetikhin Poisson bracket is already known \cite{Delduc:2012mk}.

\appendix

\section{Compact real form} \label{app: real form}

Let $F$ be a compact Lie group with Lie algebra $\f = \text{Lie}(F)$. We denote by $\f^{\CC}$ the complexification of $\f$ and fix a choice of Cartan subalgebra $\h$ with corresponding root space decomposition
\begin{equation*}
\textstyle \f^{\CC} = \h \bigoplus \big( \! \oplus_{\alpha \in \Phi} \CC E^{\alpha} \big).
\end{equation*}
Given a choice of simple roots $\alpha_i \in \Phi$, $i = 1, \ldots, n = \text{rk} \, \f^{\CC}$ we denote the pair of opposite nilpotent subalgebras as $\n^{\pm} = \oplus_{\alpha > 0} \CC E^{\pm \alpha}$ and the corresponding Borel subalgebras as $\b^{\pm} = \h \oplus \n^{\pm}$. The non-trivial Lie algebra relations in $\f^{\CC}$ read, for any roots $\alpha, \beta \in \Phi$,
\begin{equation*}
[H, E^{\alpha}] = \alpha(H) E^{\alpha}, \qquad
[E^{\alpha}, E^{- \alpha}] = H^{\alpha}, \qquad
[E^{\alpha}, E^{\beta}] = N_{\alpha, \beta} E^{\alpha + \beta}, \quad \text{if} \;\; \alpha + \beta \in \Phi
\end{equation*}
where $H^{\alpha} \in \h$ is defined for any root $\alpha \in \Phi$ in terms of the Killing form on $\f^{\CC}$ as $\kappa(H^{\alpha}, H) = \alpha(H)$. The latter induces a (positive definite) inner product on the set of roots denoted $(\alpha, \beta) = \alpha(H^{\beta})$. We have chosen the normalisation of the generators $E^{\alpha}$ so that
\begin{equation*}
\kappa(E^{\alpha}, E^{\beta}) = \delta_{\alpha, - \beta}.
\end{equation*}
Letting $H^i = H^{\alpha_i}$ for any simple root $\alpha_i$ we have
\begin{equation*}
\kappa(H^i, H^j) = \alpha_i(H^j) = (\alpha_i, \alpha_j) = B_{ij},
\end{equation*}
where $B_{ij} = d_i A_{ij}$ denotes the symmetrised Cartan matrix with $d_i = (\alpha_i, \alpha_i) / 2$. With respect to the basis $H^i$, $i = 1, \ldots, n$ and $E^{\alpha}$, $\alpha \in \Phi$ of $\f^{\CC}$, the tensor Casimir then reads
\begin{equation} \label{Casimir HE}
C_{\1\2} = \sum_{i, j = 1}^n B^{-1}_{ij} H^i \otimes H^j + \sum_{\alpha > 0} \big( E^{\alpha} \otimes E^{-\alpha} + E^{-\alpha} \otimes E^{\alpha} \big).
\end{equation}
If $\beta + \pp \alpha, \ldots, \beta, \ldots, \beta + \qq \alpha$ denotes the
$\alpha$-string through $\beta$, where $\pp \leq 0$ and $\qq \geq 0$, then with the
above conventions one may show that
\begin{equation} \label{Npq}
N_{\alpha, \beta}^2 = \qq (1 - \pp) \frac{(\alpha, \alpha)}{2}, \qquad
\frac{2 (\beta, \alpha)}{(\alpha, \alpha)} = - (\pp + \qq).
\end{equation}
In particular, the structure constants $N_{\alpha, \beta}$ are all real.

The real Lie algebra $\f$ is recovered from its complexification $\f^{\CC}$ as the fixed point set of a certain anti-linear involutive automorphism $\tau$, namely such that
\begin{equation*}
\tau(\lambda X + \mu Y) = \overline{\lambda} \, \tau(X) + \overline{\mu} \, \tau(Y), \qquad
\tau^2 = 1, \qquad \tau( [X, Y] ) = [\tau(X), \tau(Y)],
\end{equation*}
for any $X, Y \in \f^{\CC}$ and $\lambda, \mu \in \CC$. It is convenient to define $\tau(X) = - X^{\dag}$ in terms of an anti-linear involutive anti-automorphism $X \mapsto X^{\dag}$ with the properties
\begin{equation*}
(\lambda X + \mu Y)^{\dag} = \overline{\lambda} \, X^{\dag} + \overline{\mu} \, Y^{\dag}, \qquad (X^{\dag})^{\dag} = X, \qquad [X, Y]^{\dag} = [Y^{\dag}, X^{\dag}].
\end{equation*}
In the case of the compact real form we define the latter on the basis $H^i, E^{\alpha}$ as
\begin{equation} \label{herm conj}
(H^i)^{\dag} = H^i, \qquad (E^{\alpha})^{\dag} = E^{-\alpha}.
\end{equation}
We then have by definition $\f = \{ X \in \f^{\CC} \,|\, \tau(X) = X \}$.
A basis over $\mathbb{R}$ for the compact real form $\f$ is then given by
\begin{equation} \label{compact generators}
T^i = i H^i, \qquad B^{\alpha} = \frac{i}{\sqrt{2}} (E^{\alpha} + E^{- \alpha}), \qquad C^{\alpha} = \frac{1}{\sqrt{2}} (E^{\alpha} - E^{- \alpha}).
\end{equation}
With respect to these generators the Killing form reads
\begin{equation} \label{Killing form compact}
\kappa(T^i, T^j) = - B_{ij}, \qquad
\kappa(B^{\alpha}, B^{\beta}) = - \delta_{\alpha, - \beta}, \qquad
\kappa(C^{\alpha}, C^{\beta}) = - \delta_{\alpha, - \beta}
\end{equation}
so that the tensor Casimir may be expressed as
\begin{equation} \label{Casimir TBC}
C_{\1\2} = - \sum_{i, j = 1}^n B^{-1}_{ij} T^i \otimes T^j - \sum_{\alpha > 0} \big( B^{\alpha} \otimes B^{\alpha} + C^{\alpha} \otimes C^{\alpha} \big).
\end{equation}

\paragraph{Iwasawa decomposition.} Let $\h_0$ denote the linear span over $\mathbb{R}$ of the set of Cartan generators $H^i$, $i = 1, \ldots, n$. Then the lower Borel subalgebra $\b^- \in \f^{\CC}$ is contained in $\f \oplus \h_0 \oplus \n^+$. Indeed, any element in $\b^-$ takes the form $X + h$ where $X = \sum_{\alpha > 0} x_{\alpha} E^{- \alpha} \in \n^-$ and $h = \sum_{i = 1}^n a_i H^i$ for some $x_{\alpha}, a_i \in \CC$. It then follows using \eqref{herm conj} that $X^{\dag} = \sum_{\alpha > 0} \overline{x}_{\alpha} E^{\alpha} \in \n^+$ and hence
\begin{equation*}
X + h = \big( (X + \ha h) - (X + \ha h)^{\dag} \big) + \ha (h + h^{\dag}) + X^{\dag} \in \f \oplus \h_0 \oplus \n^+.
\end{equation*}
In particular, using the decomposition $\f^{\CC} = \b^- \oplus \n^+$ it follows that
\begin{equation} \label{Iwasawa}
\f^{\CC} = \f \oplus \h_0 \oplus \n^+.
\end{equation}
This is known as the Iwasawa decomposition of the complex Lie algebra $\f^{\CC}$.

\section{$q$-Poisson-Serre relations} \label{app: q-Serre}

In this appendix we prove the $q$-Poisson-Serre relations \eqref{q-Serre}. To do this we will define charges associated also with non-simple roots $\alpha \in \Phi^+$. This in turn requires choosing a normal ordering on the set of positive roots $\Phi^+$ of $\f^{\CC}$ (see for instance \cite{Asherova_1979, Tolstoy_1989a, Khoroshkin_1991}), namely such that if $\alpha < \beta$ and $\alpha + \beta$ is a root then $\alpha < \alpha + \beta < \beta$.
Given such a choice of ordering, we write the nilpotent part of the monodromy matrix
\eqref{A factor Cartan a} as follows
\begin{equation*}
P \overleftarrow{\exp} \biggl[ \gamma \, \sum_{\alpha > 0} \int_{- \infty}^{\infty} d\sigma \, \QQ^E_{\alpha}(\sigma) E^{\alpha} \biggr]
= \sideset{}{^<}\prod_{\alpha > 0} \exp \biggl( \gamma \, \int_{- \infty}^{\infty} d\sigma \, \mathfrak{Q}^E_{\alpha}(\sigma) E^{\alpha} \biggr),
\end{equation*}
where the superscript $<$ on the product indicates the use of normal ordering on the positive roots.
Note that the normal ordering only defines a partial ordering on the set of positive roots. However, whenever two roots $\alpha$ and $\beta$ are not ordered this implies that $\alpha + \beta$ is not a root. It follows that the corresponding generators $E^{\alpha}$ and $E^{\beta}$ commute and therefore their relative order in the above product is irrelevant.

Let $\alpha_i$, $\alpha_j$ be simple roots and consider the collection of roots
$\alpha$ belonging to the $\alpha_i$-string through
$\alpha_j$, namely $\alpha_j, \alpha_j + \alpha_i, \ldots, \alpha_j + \qq \, \alpha_i$
for some $\qq \geq 0$ such that $\alpha_j + (\qq+1) \alpha_i$ is not a root.
It is easy to see that for the simple root $\alpha_j$ we have
\begin{equation*}
\mathfrak{Q}^E_{\alpha_j}(\sigma) = \QQ^E_{\alpha_j}(\sigma).
\end{equation*}
Next, consider the sum $\alpha_j + \alpha_i$. Assuming this is a root, which is the case if
$\qq \geq 1$, we must have either $\alpha_j < \alpha_i$ or
$\alpha_i < \alpha_j$. It will be more convenient to work with a normal ordering
such that $\alpha_j < \alpha_i$. In this case it follows that the roots of the $\alpha_i$-string
through $\alpha_j$ are ordered as
\begin{equation*}
\alpha_j < \alpha_j + \alpha_i < \alpha_j + 2 \alpha_i < \ldots < \alpha_j + \qq \,
\alpha_i < \alpha_i.
\end{equation*}
The charge density corresponding to the sum of simple roots $\alpha_j + \alpha_i$ is found to be
\begin{equation} \label{Q ai+aj}
\mathfrak{Q}^E_{\alpha_j + \alpha_i}(\sigma) = \QQ^E_{\alpha_j + \alpha_i}(\sigma) - \gamma \, N_{\alpha_j, \alpha_i} \QQ^E_{\alpha_i}(\sigma) \int_{- \infty}^{\sigma} d\sigma' \, \QQ^E_{\alpha_j}(\sigma').
\end{equation}
More generally, the charge density $\mathfrak{Q}^E_{\alpha_j + r \alpha_i}(\sigma)$ associated
with the root $\alpha_j + r \alpha_i$ with $0 < r \leq \qq$ may be expressed recursively in terms of the preceding charge density $\mathfrak{Q}^E_{\alpha_j + (r-1) \alpha_i}(\sigma)$ as follows
\begin{equation} \label{Q ai+r aj}
\mathfrak{Q}^E_{\alpha_j + r \alpha_i}(\sigma) = \QQ^E_{\alpha_j + r \alpha_i}(\sigma) - \gamma \, N_{\alpha_j + (r-1) \alpha_i, \alpha_i} \QQ^E_{\alpha_i}(\sigma) \int_{- \infty}^{\sigma} d\sigma' \, \mathfrak{Q}^E_{\alpha_j + (r-1) \alpha_i}(\sigma').
\end{equation}
Finally, recalling the notation \eqref{Di} we define the charges corresponding to each root $\alpha_j + r \alpha_i$ as
\begin{equation} \label{Q roots def}
Q^E_{\alpha_j + r \alpha_i} = D_j D_i^r \int_{-\infty}^{\infty} d\sigma \, \mathfrak{Q}^E_{\alpha_j + r \alpha_i}(\sigma),
\end{equation}
so that in the case $r = 0$ this definition agrees with \eqref{simple charges}.

In the remainder of this appendix we will prove that the generators defined in
\eqref{Q roots def} satisfy the following Poisson algebra with respect to the
$q$-Poisson bracket introduced in \eqref{q-bracket}, for $r \leq \qq$,
\begin{equation} \label{Cartan-Weyl}
\big\{ Q^E_{\alpha_i}, Q^E_{\alpha_j + r \alpha_i} \big\}_{q \, \epsilon} = 2 i N_{\alpha_j + r \alpha_i, \alpha_i} Q^E_{\alpha_j + (r + 1) \alpha_i}.
\end{equation}
Since $\alpha_j + (\qq+1) \alpha_i$ is not a root by definition of $\qq$
we have that $N_{\alpha_j + \qq \alpha_i, \alpha_i} = 0$. It therefore follows
from \eqref{Cartan-Weyl} that
\begin{equation*}
\big\{ \underbrace{Q^E_{\alpha_i}, \big\{ Q^E_{\alpha_i}, \ldots
\big\{ Q^E_{\alpha_i}}_{\qq + 1 \; \text{times}}, Q^E_{\alpha_j} \big\}_{q \, \epsilon} \ldots \big\}_{q \, \epsilon} \big\}_{q \, \epsilon} = 0,
\end{equation*}
which is nothing but the $q$-Poisson-Serre relation \eqref{q-Serre} since $\qq = - A_{ij}$.
In fact, to establish the $q$-Poisson-Serre relations for classical Lie algebras it suffices to
show that \eqref{Cartan-Weyl} holds with $r \leq 2$ since for every pair of simple roots
$\alpha_i$ and $\alpha_j$, the $\alpha_i$-string through $\alpha_j$ has at most $\qq = 2$.

\paragraph{Case $r=0$.} We begin by proving the relation \eqref{Cartan-Weyl} in the case $r = 0$. Comparing coefficients of $E^{\beta}$ on both sides of the second relation in \eqref{hX eX PB} yields
\begin{equation*}
\{ e_{\alpha}(\sigma), e_{\beta}(\sigma') \}_{\epsilon} = 2 i N_{\beta, \alpha}
e_{\alpha + \beta}(\sigma) \delta_{\sigma \sigma'}, \qquad \text{if} \;\; \alpha + \beta \in \Phi.
\end{equation*}
Using the definition \eqref{july7d} of $\QQ^E_{\alpha}$ this then leads to
\begin{equation} \label{PB JJ}
\{ \QQ^E_{\alpha}(\sigma), \QQ^E_{\beta}(\sigma') \}_{\epsilon} = 2 i N_{\beta, \alpha} \QQ^E_{\alpha + \beta}(\sigma) \delta_{\sigma \sigma'} + i \gamma \, (\alpha, \beta) \QQ^E_{\alpha}(\sigma) \QQ^E_{\beta}(\sigma') \epsilon_{\sigma \sigma'}.
\end{equation}
Introducing the Heaviside step function $\theta_{\sigma \sigma'} = \ha (\epsilon_{\sigma \sigma'} + 1)$ we may rewrite this as
\begin{equation*}
\{ \QQ^E_{\alpha}(\sigma), \QQ^E_{\beta}(\sigma') \}_{\epsilon} + i \gamma \, (\alpha, \beta) \QQ^E_{\alpha}(\sigma) \QQ^E_{\beta}(\sigma') = 2 i \bigl( N_{\beta, \alpha} \QQ^E_{\alpha + \beta}(\sigma) \delta_{\sigma \sigma'} + \gamma \, (\alpha, \beta) \QQ^E_{\alpha}(\sigma) \QQ^E_{\beta}(\sigma') \theta_{\sigma \sigma'} \bigr).
\end{equation*}
In terms of the $q$-Poisson bracket introduced in \eqref{q-bracket}, it now follows from the above in the case $\alpha = \alpha_i$ and $\beta = \alpha_j$ that
\begin{align*}
\big\{ Q^E_{\alpha_i}, Q^E_{\alpha_j} \big\}_{q \, \epsilon} &= \{ Q^E_{\alpha_i}, Q^E_{\alpha_j} \}_{\epsilon} + i \gamma \, (\alpha_i, \alpha_j) Q^E_{\alpha_i} Q^E_{\alpha_j}\\
&= 2 i D_i D_j \biggl( N_{\alpha_j, \alpha_i} \int_{- \infty}^{\infty} d\sigma \, \QQ^E_{\alpha_j + \alpha_i}(\sigma) + \gamma \, (\alpha_j, \alpha_i) \int_{-\infty}^{\infty} d\sigma \, \QQ^E_{\alpha_i}(\sigma) \int_{-\infty}^{\sigma} d\sigma' \, \QQ^E_{\alpha_j}(\sigma') \biggr).
\end{align*}
Now using \eqref{Npq} with $\alpha = \alpha_i$ and $\beta = \alpha_j$, since
$\alpha_j - \alpha_i$ is not a root we have $\pp = 0$ from which we deduce that $N^2_{\alpha_j, \alpha_i} = - (\alpha_j, \alpha_i)$.
Hence we deduce using the definitions \eqref{Q ai+aj} and \eqref{Q roots def} that
\begin{equation*}
\big\{ Q^E_{\alpha_i}, Q^E_{\alpha_j} \big\}_{q \, \epsilon} = 2 i N_{\alpha_j, \alpha_i} Q^E_{\alpha_j + \alpha_i}.
\end{equation*}

\paragraph{Cases $r=1$ and $r=2$.}

The relation \eqref{Cartan-Weyl} in the cases $r = 1$ and $r = 2$ follows in a similar way. For instance, starting from the definition \eqref{Q ai+aj} and the relation \eqref{PB JJ} one can show that
\begin{align*}
\{ \QQ^E_{\alpha_i}(\sigma), \, &\mathfrak{Q}^E_{\alpha_j + \alpha_i}(\sigma') \}_{\epsilon} + i \gamma \, (\alpha_i, \alpha_j + \alpha_i) \QQ^E_{\alpha_i}(\sigma) \mathfrak{Q}^E_{\alpha_j + \alpha_i}(\sigma')\\
&= 2 i \Bigl( N_{\alpha_j + \alpha_i, \alpha_i} \QQ^E_{\alpha_j + 2 \alpha_i}(\sigma) \delta_{\sigma \sigma'}\\
&\qquad + \gamma \, (\alpha_i, \alpha_j + \alpha_i) \QQ^E_{\alpha_i}(\sigma) \mathfrak{Q}^E_{\alpha_j + \alpha_i}(\sigma') \theta_{\sigma \sigma'} + \gamma \, (\alpha_i, \alpha_j) \QQ^E_{\alpha_i}(\sigma') \mathfrak{Q}^E_{\alpha_j + \alpha_i}(\sigma) \theta_{\sigma' \sigma} \Bigr).
\end{align*}
Taking the integral over $\sigma$ and $\sigma'$ then yields the desired relation \eqref{Cartan-Weyl} in the case $r = 1$, namely
\begin{equation*}
\big\{ Q^E_{\alpha_i}, Q^E_{\alpha_j + \alpha_i} \big\}_{q \, \epsilon} = 2 i N_{\alpha_j + \alpha_i, \alpha_i} Q^E_{\alpha_j + 2 \alpha_i}.
\end{equation*}
In deriving these results we make use of the following useful identities, valid for any $0 \leq r \leq \qq$,
\begin{gather*}
N^2_{\alpha_j + r \alpha_i, \alpha_i} = - \bigl( (r+1) \alpha_j + \mbox{\small $\frac{r (r+1)}{2}$} \alpha_i, \alpha_i \bigr),\\
- N^2_{\alpha_j + (r-1) \alpha_i, \alpha_i} + (\alpha_j + r \alpha_i, \alpha_i) = - N^2_{\alpha_j + r \alpha_i, \alpha_i}.
\end{gather*}

Finally, in the case $r = 2$, a lengthy calculation leads to the following
\begin{align*}
\{ \QQ^E_{\alpha_i}(\sigma), \, &\mathfrak{Q}^E_{\alpha_j + 2 \alpha_i}(\sigma') \}_{\epsilon} + i \gamma \, (\alpha_i, \alpha_j + 2 \alpha_i) \QQ^E_{\alpha_i}(\sigma) \mathfrak{Q}^E_{\alpha_j + 2 \alpha_i}(\sigma')\\
&= 2 i \biggl( N_{\alpha_j + 2 \alpha_i, \alpha_i} \QQ^E_{\alpha_j + 3 \alpha_i}(\sigma) \delta_{\sigma \sigma'}\\
&\qquad\qquad + \gamma \, (\alpha_i, \alpha_j + 2 \alpha_i) \QQ^E_{\alpha_i}(\sigma) \QQ^E_{\alpha_j + 2 \alpha_i}(\sigma') \theta_{\sigma \sigma'} - \gamma \, N^2_{\alpha_j + \alpha_i, \alpha_i} \QQ^E_{\alpha_i}(\sigma') \QQ^E_{\alpha_j + 2 \alpha_i}(\sigma) \theta_{\sigma' \sigma}\\
&\qquad\qquad - \gamma^2 N_{\alpha_j + \alpha_i, \alpha_i} \biggl( (\alpha_i, \alpha_j + \alpha_i) \QQ^E_{\alpha_i}(\sigma) \QQ^E_{\alpha_i}(\sigma') \int_{- \infty}^{\sigma'} d\sigma'' \mathfrak{Q}^E_{\alpha_j + \alpha_i}(\sigma'') \theta_{\sigma \sigma'}\\
&\qquad\qquad\qquad\qquad\qquad + (\alpha_i, \alpha_j + \alpha_i) \QQ^E_{\alpha_i}(\sigma) \QQ^E_{\alpha_i}(\sigma') \int_{- \infty}^{\sigma} d\sigma'' \mathfrak{Q}^E_{\alpha_j + \alpha_i}(\sigma'') \theta_{\sigma' \sigma}\\
&\qquad\qquad\qquad\qquad\qquad + (\alpha_i, \alpha_i) \QQ^E_{\alpha_i}(\sigma) \QQ^E_{\alpha_i}(\sigma') \int_{- \infty}^{\sigma'} d\sigma'' \mathfrak{Q}^E_{\alpha_j + \alpha_i}(\sigma'') \theta_{\sigma \sigma'}\\
&\qquad\qquad\qquad\qquad\qquad + (\alpha_i, \alpha_j ) \QQ^E_{\alpha_i}(\sigma') \mathfrak{Q}^E_{\alpha_j + \alpha_i}(\sigma) \int_{- \infty}^{\sigma'} d\sigma'' \QQ^E_{\alpha_i}(\sigma'') \theta_{\sigma' \sigma}\\
&\qquad\qquad\qquad\qquad\qquad - (\alpha_i, \alpha_j ) \QQ^E_{\alpha_i}(\sigma') \mathfrak{Q}^E_{\alpha_j + \alpha_i}(\sigma) \int_{- \infty}^{\sigma} d\sigma'' \QQ^E_{\alpha_i}(\sigma'') \theta_{\sigma' \sigma}
\biggr) \biggr).
\end{align*}
After taking the integral over $\sigma$ and $\sigma'$ we obtain the sought after relation \eqref{Cartan-Weyl} with $r = 2$, namely
\begin{equation*}
\big\{ Q^E_{\alpha_i}, Q^E_{\alpha_j + 2 \alpha_i} \big\}_{q \, \epsilon} = 2 i N_{\alpha_j + 2 \alpha_i, \alpha_i} Q^E_{\alpha_j + 3 \alpha_i}.
\end{equation*}

\section{Modified classical Yang-Baxter equation} \label{app: mCYBE}

The modified classical Yang-Baxter equation \eqref{mCYBE non-split} satisfied by the $R$-matrix \eqref{R def} in the present article (see also \cite{Klimcik:2008eq}) is slightly different from the one which appeared in \cite{Delduc:2012qb}. The general form of this equation over a real Lie algebra $\f$ reads
\begin{equation} \label{mCYBE}
[RX, RY] - R\big( [RX, Y] + [X, RY] \big) = - \omega [X, Y],
\end{equation}
for some real parameter $\omega \in \mathbb{R}$. Of course, by rescaling the linear map $R \in \text{End}\, \f$ by $1/\sqrt{|\omega|}$ we may restrict attention to the cases $\omega = \pm 1$.
The $R$-matrices discussed in \cite{Delduc:2012qb} are solutions of this equation with $\omega = 1$, sometimes referred to as the `split case'. However, the $R$-matrix \eqref{R def} used here and in \cite{Klimcik:2008eq} is a solution of this equation with $\omega = -1$, referred to as the `non-split case'.

In either case, the modified classical Yang-Baxter equation \eqref{mCYBE} may be rewritten as
\begin{equation} \label{mCYBE hom}
(R \pm \sqrt{\omega}) \big( [X, Y]_R \big) = \big[ (R \pm \sqrt{\omega}) X, (R \pm \sqrt{\omega}) Y \big],
\end{equation}
where $[X, Y]_R \coloneqq [RX, Y] + [X, RY]$ defines a second Lie bracket on $\f$ by virtue of \eqref{mCYBE}. In the split case, this implies that the linear maps $R_{\pm} \coloneqq R \pm 1$ are both Lie algebra homomorphisms $\f_R \to \f$ where $\f_R$ is the vector space $\f$ equipped with the Lie bracket $[\cdot, \cdot]_R$. In the non-split case, however, things are a little more subtle. Since $\sqrt{\omega} = i$, we see that the linear maps $R \pm i$ are still Lie algebra homomorphisms by \eqref{mCYBE hom} but now from $\f_R \to \f^{\CC}$. Recall that in the split case ($\omega = 1$), the pair of maps $R_{\pm}$ can be used to define an embedding $\f_R \to \f \oplus \f$ so that $\f_R$ may be regarded as a subalgebra of the double $\f \oplus \f$. In the present non-split case ($\omega = -1$), however, the map $R - i$ alone defines an embedding of the real Lie algebra $\f_R$ into the complexification $\f^{\CC}$.

\section{Deformed Poisson bracket for coset $\sigma$-model} \label{app: int alg}

The deformed Poisson bracket \eqref{deformed PB coset}, when expressed in terms of the graded components of the fields $A$ and $\Pi$, takes the following form
\begin{align*}
\{ A^{(0)}_{\1}(\sigma), A^{(0)}_{\2}(\sigma') \}_{\epsilon} &= - \epsilon^2 \big[ C^{(00)}_{\1\2}, 2 A^{(0)}_{\2}(\sigma) + \Pi^{(0)}_{\2}(\sigma) \big] \delta_{\sigma \sigma'} + 2 \epsilon^2 C^{(00)}_{\1\2} \delta'_{\sigma \sigma'},\\
\{ A^{(0)}_{\1}(\sigma), A^{(1)}_{\2}(\sigma') \}_{\epsilon} &= - \epsilon^2 \big[ C^{(00)}_{\1\2}, A^{(1)}_{\2}(\sigma) + \Pi^{(1)}_{\2}(\sigma) \big] \delta_{\sigma \sigma'},\\
\{ A^{(1)}_{\1}(\sigma), A^{(1)}_{\2}(\sigma') \}_{\epsilon} &= - \epsilon^2 \big[ C^{(11)}_{\1\2}, \Pi^{(0)}_{\2}(\sigma) \big] \delta_{\sigma \sigma'},\\
\{ A^{(0)}_{\1}(\sigma), \Pi^{(0)}_{\2}(\sigma') \}_{\epsilon} &= \big[ C^{(00)}_{\1\2}, A^{(0)}_{\2}(\sigma) \big] \delta_{\sigma \sigma'} - C^{(00)}_{\1\2} \delta'_{\sigma \sigma'},\\
\{ A^{(0)}_{\1}(\sigma), \Pi^{(1)}_{\2}(\sigma') \}_{\epsilon} &= (1 - \epsilon^2) \big[ C^{(00)}_{\1\2}, A^{(1)}_{\2}(\sigma) \big] \delta_{\sigma \sigma'} - \epsilon^2 \big[ C^{(00)}_{\1\2}, \Pi^{(1)}_{\2}(\sigma) \big] \delta_{\sigma \sigma'},\\
\{ A^{(1)}_{\1}(\sigma), \Pi^{(0)}_{\2}(\sigma') \}_{\epsilon} &= \big[ C^{(11)}_{\1\2}, A^{(1)}_{\2}(\sigma) \big] \delta_{\sigma \sigma'},\\
\{ A^{(1)}_{\1}(\sigma), \Pi^{(1)}_{\2}(\sigma') \}_{\epsilon} &= \big[ C^{(11)}_{\1\2}, A^{(0)}_{\2}(\sigma) \big] \delta_{\sigma \sigma'} + \epsilon^2 \big[ C^{(11)}_{\1\2}, \Pi^{(0)}_{\2}(\sigma) \big] \delta_{\sigma \sigma'} - C^{(11)}_{\1\2} \delta'_{\sigma \sigma'},\\
\{ \Pi^{(0)}_{\1}(\sigma), \Pi^{(0)}_{\2}(\sigma') \}_{\epsilon} &= \big[ C^{(00)}_{\1\2}, \Pi^{(0)}_{\2}(\sigma) \big] \delta_{\sigma \sigma'},\\
\{ \Pi^{(0)}_{\1}(\sigma), \Pi^{(1)}_{\2}(\sigma') \}_{\epsilon} &= \big[ C^{(00)}_{\1\2}, \Pi^{(1)}_{\2}(\sigma) \big] \delta_{\sigma \sigma'},\\
\{ \Pi^{(1)}_{\1}(\sigma), \Pi^{(1)}_{\2}(\sigma') \}_{\epsilon} &= (1 - \epsilon^2) \big[ C^{(11)}_{\1\2}, \Pi^{(0)}_{\2}(\sigma) \big] \delta_{\sigma \sigma'}.
\end{align*}

\providecommand{\href}[2]{#2}\begingroup\raggedright\endgroup

\end{document}

%% file: poles_coset.eps_tex
\begingroup%
  \makeatletter%
  \providecommand\color[2][]{%
    \errmessage{(Inkscape) Color is used for the text in Inkscape, but the package 'color.sty' is not loaded}%
    \renewcommand\color[2][]{}%
  }%
  \providecommand\transparent[1]{%
    \errmessage{(Inkscape) Transparency is used (non-zero) for the text in Inkscape, but the package 'transparent.sty' is not loaded}%
    \renewcommand\transparent[1]{}%
  }%
  \providecommand\rotatebox[2]{#2}%
  \ifx\svgwidth\undefined%
    \setlength{\unitlength}{355.85bp}%
    \ifx\svgscale\undefined%
      \relax%
    \else%
      \setlength{\unitlength}{\unitlength * \real{\svgscale}}%
    \fi%
  \else%
    \setlength{\unitlength}{\svgwidth}%
  \fi%
  \global\let\svgwidth\undefined%
  \global\let\svgscale\undefined%
  \makeatother%
  \begin{picture}(1,0.62638752)%
    \put(0,0){\includegraphics[width=\unitlength]{poles_coset.eps}}%
    \put(0.71871124,0.46019551){\color[rgb]{0,0,0}\makebox(0,0)[lb]{\smash{$\lambda_+$}}}%
    \put(0.23006465,0.15144828){\color[rgb]{0,0,0}\makebox(0,0)[lb]{\smash{$\lambda_-$}}}%
    \put(0.75934883,0.32885924){\color[rgb]{0,0,0}\makebox(0,0)[lb]{\smash{$1$}}}%
    \put(0.17282559,0.32966209){\color[rgb]{0,0,0}\makebox(0,0)[lb]{\smash{$-1$}}}%
    \put(0.48636065,0.59702993){\color[rgb]{0,0,0}\makebox(0,0)[lb]{\smash{$i$}}}%
    \put(0.46949954,0.01411975){\color[rgb]{0,0,0}\makebox(0,0)[lb]{\smash{$-i$}}}%
    \put(0.71871124,0.14866775){\color[rgb]{0,0,0}\makebox(0,0)[lb]{\smash{$\lambda_+^{-1}$}}}%
    \put(0.22140148,0.46013445){\color[rgb]{0,0,0}\makebox(0,0)[lb]{\smash{$\lambda_-^{-1}$}}}%
  \end{picture}%
\endgroup%